\documentclass[sigconf]{acmart} 

\AtBeginDocument{%
  }

\setcopyright{acmlicensed}
\copyrightyear{2018}
\acmYear{2018}
\acmDOI{XXXXXXX.XXXXXXX}
\acmConference[Conference acronym 'XX]{Make sure to enter the correct
  conference title from your rights confirmation email}{June 03--05,
  2018}{Woodstock, NY}
\acmISBN{978-1-4503-XXXX-X/2018/06}

\usepackage{tikz}
\usepackage{amsmath}
\usepackage{url}
\usepackage{multirow} 
\usepackage{hyperref}
\usepackage{tabularx}
\usepackage{multirow}
\usepackage{ragged2e} 
\usepackage{booktabs} 
\usepackage{graphicx}
\usepackage{arydshln}
\usepackage{array}
\usepackage{tabularray}
\usepackage{enumitem}
\usepackage{makecell}
\usepackage{caption}
\usepackage{stfloats}  
\usepackage{mdframed}
\makeatletter
\@ifundefined{Bbbk}{}{%
  
}
\makeatother
\usepackage{amssymb}

\usepackage{filecontents}
\begin{document}

\title{\textit{ImmuniFraug}: A Metacognitive Intervention Anti-Fraud Approach to Enhance Undergraduate Students' Cyber Fraud Awareness}

\renewcommand{\shorttitle}{ImmuniFraug}

\author{Xiangzhe Yuan}
\affiliation{%
  \department{Department of Computer Science}
  \institution{The University of Iowa}
  \city{Iowa City}
  \country{America}
}
\email{xiangzhe-yuan@uiowa.edu}

\author{Jiajun Wang}
\affiliation{%
  \department{School of Information Systems and Technology Management}
  \institution{University of New South Wales}
  \city{Sydney}
  \country{Australia}
}
\email{jiajun.wang5@student.unsw.edu.au}

\author{Huanchen Wang}
\affiliation{%
  \department{Department of Computer Science}
  \institution{City University of Hong Kong}
  \city{Hong Kong SAR}
  \country{China}
}
\email{huancwang2-c@my.cityu.edu.hk}

\author{Qian Wan}
\affiliation{%
  \department{Department of Computer Science}
  \institution{City University of Hong Kong}
  \city{Hong Kong SAR}
  \country{China}
}
\email{qianwan3-c@my.cityu.edu.hk}

\author{Siying Hu}
\affiliation{%
  \department{Department of Computer Science}
  \institution{University of Queensland}
  \city{Brisbane}
  \country{Australia}
}
\email{siying.hu@uqconnect.edu.au}

\begin{abstract}
Cyber fraud now constitutes over half of criminal cases in China, with undergraduate students experiencing a disproportionate rise in victimization. Traditional anti-fraud training remains predominantly passive, yielding limited engagement and retention. This paper introduces \textit{ImmuniFraug}, a Large Language Model (LLM)-based metacognitive intervention that delivers immersive, multimodal fraud simulations integrating text, voice, and visual avatars across ten prevalent fraud types. Each scenario is designed to replicate real-world persuasion tactics and psychological pressure, while post-interaction debriefs provide grounded feedback in protection motivation theory and reflective prompts to reinforce learning. In a controlled study with 846 Chinese undergraduates, \textit{ImmuniFraug} was compared to official text-based materials. Linear Mixed-Effects Modeling (LMEM) reveals that the interactive intervention significantly improved fraud awareness ($p = 0.026$), successfully providing incremental learning value even when controlling for participants' extensive prior exposure to anti-fraud education, alongside high narrative immersion (M = 56.95/77). Thematic analysis of interviews revealed key effectiveness factors: perceived realism, adaptive deception, enforced time pressure, emotional manipulation awareness, and enhanced self-efficacy. Findings demonstrate that by shifting the focus from passive knowledge acquisition to active metacognitive engagement, LLM-based simulations offer a scalable and ecologically valid new paradigm for anti-fraud training and fostering fraud resilience.
\end{abstract}

\ccsdesc[500]{Security and privacy~Usability in security and privacy}
\ccsdesc[500]{Human-centered computing~Empirical studies in HCI}

\keywords{Cyber Fraud, LLM, Metacognitive Intervention, Multimodal Interaction, Anti-Fraud}

\received{20 February 2007}
\received[revised]{12 March 2009}
\received[accepted]{5 June 2009}

\maketitle

Cyber fraud has become a pervasive threat, with cyber fraud now accounting for over half of all criminal cases in China \cite{Sanlian2024}. Recent statistics show that scam incidents among Chinese undergraduate students have been rising at an alarming 23.7\% annual rate (2020–2023), far outpacing the general population’s 15.2\% growth \cite{He2025}. This trend, coupled with high-profile losses in campus fraud cases, underscores the urgent need for more effective anti-fraud education for young adults. Although universities and law enforcement agencies have organized numerous anti-fraud awareness campaigns, lectures, and training sessions, scam cases continue to occur frequently on campuses \cite{CCTV2023}. 
Traditional ``tell-only'' training approaches – passive lectures, pamphlets, or videos, often suffer from low student engagement and limited retention of security knowledge \cite{ahmed2024enhancing,rikkers2025story}. There is a clear need for more engaging and immersive learning methods that can actively involve undergraduate students and reinforce prudent behaviors.

To this end a unique line of research has emerged with an emphasis on metacognition --- the psychological ability to monitor and control one’s thoughts and behavior~\cite{ackerman2017meta,fiedler2019metacognition,norman2019metacognition}. In addressing security risks, metacognitve interventions prompts students to reflect on their own thought processes while making decisions. Prior research in security education has emphasized that training approaches which actively stimulate student to monitor, evaluate, and adjust their cognitive strategies can lead to more robust and transferable security behaviors ~\cite{canfield2019better,zumbach2020cognitive,embrey2023many}. With the recent advancements in large language models (LLMs), emerging research indicates that LLMs are now capable of simulating immersive and dynamic game environments, further enhancing user experience~\cite{wang2024survey,li2024more}. By integrating reflective questioning and real-time feedback within interactive narratives, LLM-powered systems can effectively foster undergraduates’ metacognitive awareness, which is essential for building enduring fraud resilience.

In this paper, we present \textbf{\textit{ImmuniFraug}}, an LLM-based metacognitive intervention framework designed to help undergraduate students better understand and engage with cyber fraud scenarios, thereby raising their awareness and resilience against online scams. The intervention is delivered through interactive fraud simulation conversations, in which undergraduate students participate in simulated scam conversations that prompt reflection and decision-making. Building on findings by Yuan et al.~\cite{yuan2025day}, which suggest that undergraduates prefer multimodal interactions over text-only LLM-based interactive fiction, \textit{ImmuniFraug} integrates text, voice, and visual elements to provide a multimodal interactive experience. Through speech recognition and text-to-speech, the scammer character can speak with a human voice and respond adaptively to the user’s replies – from friendly persuasion to urgent threats – as the narrative unfolds. By hearing the scammer’s voice and seeing their avatar, undergraduates are meant to feel as if they are truly ``on the call'' or chatting with a scammer. Our scenario design is based on real crime data, incorporating the ten most prevalent scam typologies recently identified by the Ministry of Public Security of China \cite{MPS2024stats}. For a detailed description and categorization of the ten prevalent fraud types addressed in this study, please refer to Appendix \ref{tab:fraud_types_description}.
This paper aims to address the following research questions:

\textbf{RQ1}: To what extent does \textit{ImmuniFraug} enhance undergraduates' awareness and vigilance against cyber fraud?

\textbf{RQ2}: How do undergraduates perceive the interactive experience with \textit{ImmuniFraug}, and how does it shape their metacognitive understanding of cyber fraud?

\textbf{RQ3}:  What are the challenges and needs in using \textit{ImmuniFraug} to enhance undergraduates' cyber fraud awareness?

To investigate these questions, we conducted a controlled user study with undergraduate students ($n = 846$), comparing our immersive training approach with a traditional method—reading standard anti-fraud educational materials. One group of participants engaged with the interactive scam scenarios using \textit{ImmuniFraug}, while a control group read a scam text. We measured learning outcomes via post-training assessments, including a fraud knowledge quiz and self-reported vigilance scales. Additionally, we administered a Transportation Scale~\cite{green2000role} to the experimental group to quantify their sense of presence and emotional investment during the interactive experience. To rigorously evaluate the intervention effect, we utilized a Linear Mixed-Effects Model (LMEM)\cite{bates2015fitting} for the primary analysis, which allowed us to isolate the system's impact while controlling for participants' prior fraud exposure and demographic backgrounds, and accounting for the inherent heterogeneity across the ten distinct scam scenarios. After the sessions, to further explore the effectiveness of \textit{ImmuniFraug}, we conducted semi-structured interviews with some participants to understand their experiences, decision-making rationales, and suggestions for improvement.

Our results offer encouraging evidence that an immersive experience can significantly improve fraud awareness and preparedness among undergraduate students. In summary, this work makes three main contributions:  
\begin{itemize}
\item  We design and implement a novel multimodal anti-fraud training system that bridges cybersecurity and HCI, demonstrating how large language models and game-based learning can be harnessed to prevent cyber fraud.
\item Through the user study, we provide an empirical study showing that \textit{ImmuniFraug} leads to higher engagement and better learning outcomes than traditional training – undergraduates in the interactive condition scored higher on fraud recognition tests and reported stronger awareness in handling fraud situations.
\end{itemize} 

\section{Related Work}

\subsection{Background of Cyber Fraud in China}

In Chinese policy and media discourse, ``telecommunications and online fraud'' is a legal umbrella term that extends well beyond voice calls to encompass remote, technology-mediated swindles via channels like mobile messaging, social media, e-commerce platforms, investment applications and screen-sharing tools~\cite{zhang2023china}. China's Anti-Telecom and Online Fraud Law defines this offense in Article 2 as the act of ``swindling...property from others'' through remote, non-contact telecommunication and network technologies \cite{Gov2022}. This broad definition covers a wide range of cyber-enabled social engineering crimes. To avoid the narrow connotation of ``telecom fraud'' in English (i.e., phone-only scams), this paper uses the more general term \textbf{cyber fraud}. However, we retain the original Chinese terminology when citing specific policies or data sources that use it.


Cyber fraud has rapidly become a dominant form of crime in China, overtaking traditional property offenses. According to reports citing Chinese Police Association data, fraud constituted roughly 60\% of all criminal cases nationwide by the first half of 2024, with the vast majority classified as cyber fraud \cite{Huxiu2024}. University undergraduates have emerged as a particularly vulnerable demographic. For instance, a large multi-institution survey of nearly 2,400 Chinese undergraduates revealed that 47.4\% had direct or peer experience with fraud, and of those affected, approximately 20\% suffered financial losses exceeding 5,000 RMB (approx. \$700 USD)\cite{ThePaper2021}. These findings underscore the persistent and serious challenges in protecting university students from financial fraud \cite{sha2024campusfraud}.

\subsection{Current Landscape and Limitations of Anti-Fraud Prevention}

In law, \textbf{fraud} is defined as intentional deception for unlawful gain\footnote{\url{https://en.wikipedia.org/wiki/Fraud}}. While its forms are diverse, the proliferation of digital technology has created unprecedented vectors for malicious actors to commit fraud \cite{bolton2002statistical}.

Existing research on fraud prevention is largely split between two paradigms: automated detection and human awareness. The \textbf{fraud detection} paradigm leverages machine learning to identify fraudulent patterns in malicious data, such as financial transactions, textual content, or communications recording \cite{hernandez2024financial,lucas2020creditcardfrauddetection}. The \textbf{fraud awareness} relies on educational interventions, such as  media campaigns, seminars, or gamified platforms designed to improve scam recognition \cite{ahmed2024enhancing,rikkers2025story,kumaraguru2007getting,wen2019hack,chung2023reducing}.



\begin{figure*}[tp]
    \centering
    \includegraphics[width=\textwidth]{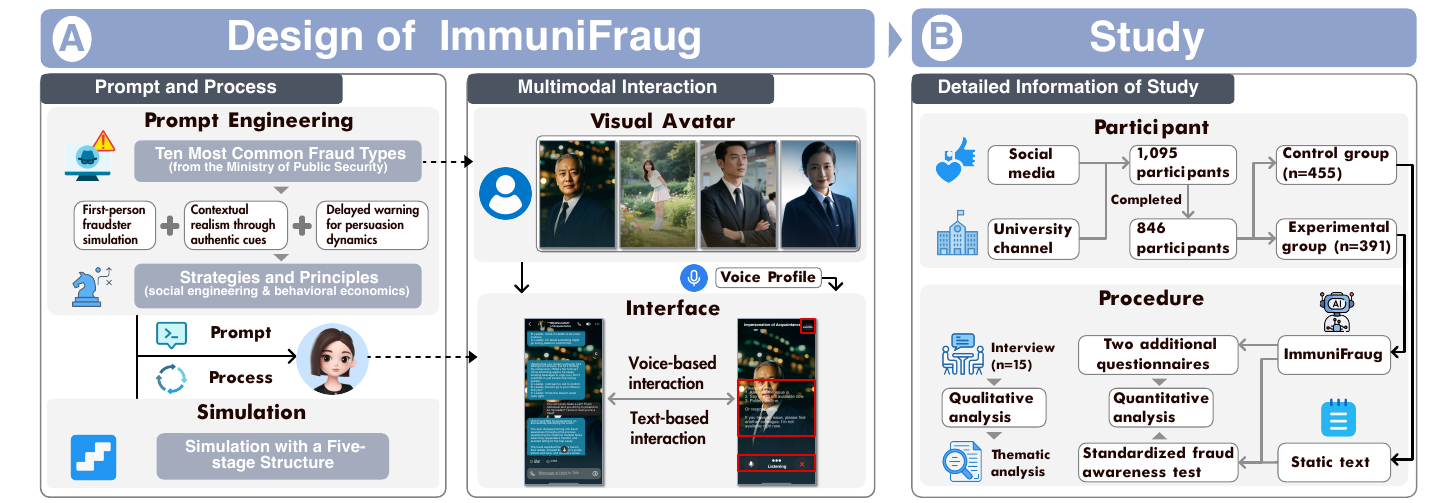}
    \caption{Workflow of the ImmuniFraug system and experimental design.}
        \label{fig:workflow}
\end{figure*}

However, while AI models drive automated detection, educational interventions remain rooted in outdated, non-interactive paradigms. These ``tell-only'' strategies, which rely on static content like lectures or videos, fail to cultivate the critical thinking and behavioral resilience necessary to navigate complex, real-world threats \cite{ahmed2024enhancing,rikkers2025story}.
More critically, the nascent research on metacognition in cybersecurity has narrowly focused on phishing, leaving a significant gap in understanding how these metacognitive processes apply to other fraud types. This limited scope restricts the generalizability and impact of interventions, which are often not adapted to the diverse psychological dynamics of other prevalent scams beyond email phishing \cite{li2021metacognitive,chen2024effects,canfield2019better}.


Our work is the first to address these gaps by leveraging Large Language Models (LLMs) for scalable, interactive fraud education. We introduce \textit{ImmuniFraug}, a framework that moves beyond the passive ``tell-only'' strategies of prior work. By engaging undergraduates in immersive, simulated scam conversational experience, \textit{ImmuniFraug} fosters the real-time reflection and decision-making skills essential for building behavioral resilience against a wide spectrum of real-world threats.

\subsection{Metacognitive Intervention Approaches in the Security Domain}

Metacognitive interventions train users to monitor, evaluate, and regulate their own reasoning during security tasks. Such approaches have shown promise across multiple domains. In an academic setting, Cano et al. found that students trained with a ``plan-monitor-reflect'' routine showed deeper learning of secure-coding principles than a control group \cite{cano2017developing}. In phishing detection, Canfield et al. revealed that users are often poorly calibrated, showing high confidence when misidentifying malicious emails; the authors suggested that metacognitive feedback on confidence could be a remedy \cite{canfield2019better}. This is supported by lab studies where prompting for self-explanation during security games enhanced knowledge acquisition \cite{zumbach2020cognitive}, and by field studies showing a strong correlation between security analysts' metacognitive awareness and their real-world performance in threat identification \cite{thangavelu2021impact}. This body of work collectively motivates the need for interventions like \textit{ImmuniFraug}, which moves beyond passive instruction to embed active reflection and confidence calibration within immersive training scenarios.

\section{Methodology}

\subsection{Design of \textit{ImmuniFraug}}

\subsubsection{Process and prompt design}


All interactive experiments were deployed on the Doubao.AI\footnote{\url{https://www.doubao.com}}, a large-scale Chinese generative AI platform. We selected Doubao because its stable Web-API, compliance with local regulations, and native Simplified Chinese interface eliminated the infrastructural and access barriers associated with international models (e.g., OpenAI's GPT), enabling broad and demographically diverse participant recruitment. The overall workflow is illustrated in Figure \ref{fig:workflow}.

We conducted prompt engineering \cite{white2023prompt,marvin2023prompt} for simulating the ten most common fraud types published by the Ministry of Public Security of China \cite{MPS2024stats}. Each simulation followed a standardized five-stage structure to ensure comparability: (1) pre-simulation disclaimer, (2) interactive fraud dialogue, (3) branching response options, (4) termination conditions, and (5) post-simulation debriefing.
During the interactive conversational simulation:
\begin{enumerate}
    \item \textit{ImmuniFraug} adopted the first-person perspective of the fraudster, customizing its identity to authentically perform the full course of the targeted scam types.
    \item The conversation incorporates credible contextual cues (e.g., order numbers, official terminology) derived from the characteristics of each scam type and real-world cases provided by the Ministry of Public Security.
    \item Obvious ``red flag'' warnings are intentionally delayed to replicate the progressive persuasion strategies characteristic of actual scams.
\end{enumerate}



At each turn, undergraduate students faced three pre-set response options and an open-text input box. The design of the options was informed by social engineering strategies that create a constrained, high-pressure decision environment to nudge victims toward compliance \cite{lopes2024common}. This pressure, often framed by scarcity or urgency (e.g., ``delayed action may affect your compensation''), disrupts deliberative processing in favor of heuristics \cite{siddiqi2022study,greavu2025exploring}. By reproducing this constrained space, we allow undergraduates to experience the psychological pressure of real scams. The open-text option restores participant agency, enabling advanced users to attempt ``off-script'' responses \cite{jalali2019decision,redmiles2018dancing}. This design operationalizes the behavioral economics principles of ``nudge'' and ``choice architecture'' in a controlled setting \cite{leonard2008richard}.


The interaction protocol was designed for a minimum of five exchanges to ensure sufficient depth, gradually leading undergraduates deeper into the trap and progressively escalating the scenario's complexity. If a undergraduate showed hesitation, \textit{ImmuniFraug} would deploy persuasive tactics to prevent premature disengagement. The simulation would conclude if the undergraduate identified the scam, refused to continue, or the maximum interaction depth was reached. The example prompt can refer Appendix~\ref{sapmpleprompt}
The simulation concludes and transitions to the debriefing phase under any of the following conditions: (1) the user explicitly refuses to continue or identifies the scam; (2) the user voluntarily terminates the session; (3) the pre-defined interaction depth (five rounds) is reached.

Research has shown that cultivating metacognitive awareness and self-efficacy can facilitate the translation of knowledge into safer behaviors ~\cite{chen2024effects,arianto2024evaluating}. Therefore, once the scam conversation ends, regardless of whether the user successfully resisted, the system transitions from the ``fraudster'' role to the debriefing stage, and prompts the user to reflect on their decision making and thought processes by offering:


\begin{itemize}
\item An assessment of the interaction's outcome, with a detailed analysis of the critical moments where deception occurred or was attempted.
\item An analysis of the trust-building techniques employed by the virtual scammer.
\item A behavioral analysis grounded in Protection Motivation Theory (PMT)~\cite{boer1996protection,marikyan2023protection}, covering threat severity, vulnerability, response efficacy, and self-efficacy.
\item Reflective prompts that guide participants to analyze their choices and consider future improvements.
\item Actionable feedback, including a performance score, a ``Fraud Prevention Keywords Reminder Card,'' and personalized advice based on the conversational recording.
\end{itemize}


This research protocol was approved by the institutional review board (IRB).  Before the simulation, a disclaimer stated that the exercise was for educational purposes, might contain persuasive or distressing language, and should not be imitated. Participants were explicitly instructed not to disclose real personal information or engage in financial transactions. An emergency ``exit'' keyword was provided to immediately terminate the simulation and access support resources. A representative conversation log is provided in Appendix~\ref{ConversationLog}

\subsubsection{Design of the multimodal interaction}





\begin{figure*}[!t]
    \centering
    \includegraphics[width=\textwidth]{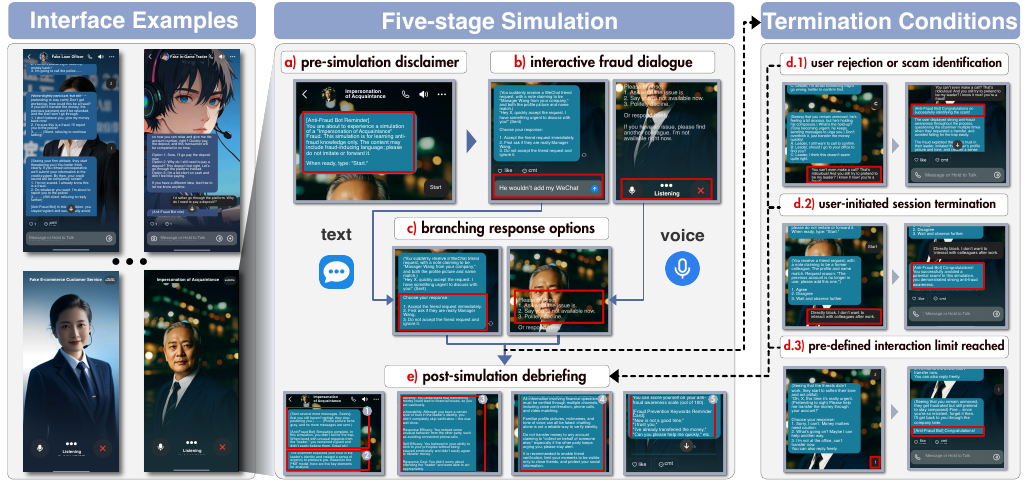}
    \caption{User interface and interactive simulation flow in \textit{ImmuniFraug}. The system integrates text, voice, and avatar-based modalities to deliver realistic fraud scenarios. \textbf{(Left)} User interface examples, where participants engage with avatars via text or voice. \textbf{(Center)} The simulation includes: (a) a pre-simulation disclaimer, (b) interactive fraud dialogue with contextual realism, and (c) branching response options guiding narrative progression. \textbf{(Right)} Sessions terminate when (d.1) the scam is rejected or identified, (d.2) users voluntarily exit, or (d.3) a predefined interaction limit is reached. Finally, (e) a post-simulation debrief provides feedback and reinforces awareness. These design choices balance immersion with controlled exposure, replicating real-world scam dynamics while safeguarding participants, and are grounded in metacognitive principles that foster awareness.}
        \label{fig:interface}
\end{figure*}

To create a highly engaging and authentic simulation, our system moves beyond text-only interaction by incorporating multimodal features enabled by the Doubao AI platform. Specifically, the system delivers responses via synthesized voice interaction and accepts voice input from users, fostering a natural and immersive conversational flow that is critical for realistic anti-fraud training. 
Grounded in research demonstrating the superiority of multimodal engagement~\cite{zhang2024unveiling}, we designed the tailored visual avatars (e.g.,Elderly leaders or young good-looking people) and voice profiles for each fraud archetype. These personas were systematically developed to align with the narratives, target demographics, and manipulative techniques of real-world scams~\cite{MPS2024stats}, with detailed specifications in Table \ref{tab:MDe}. This approach is crucial because it allows the simulation to incorporate the subtle but powerful paralinguistic cues—such as tone, pace, and emotional inflection, that are central to social engineering and absent in text-only models.
By integrating these visual and auditory elements, our system replicates the psychological pressure and nuanced persuasion of real-world threats. For instance, in addition to offering text-based simulation modes, our interface also simulates a realistic mobile phone call. Users are presented with the scammer's profile avatar and engage in a live voice conversation, effectively mirroring the high-pressure experience and psychological immediacy of an actual scam call. This provides a training experience that is not only more engaging but also more ecologically valid, equipping undergraduates with a richer understanding of how scams operate in practice. The entire interface can be seen in Figure \ref{fig:interface}

\subsection{Procedure}

To evaluate the effectiveness of \textit{ImmuniFraug}, we conducted a between-subjects experiment conducted on the Credamo online platform\footnote{\url{https://www.credamo.com/}}. The design featured two primary conditions (experimental vs. control) across ten distinct fraud scenarios. This design included two independent variables: 1) \textbf{Interactive Modality}, which constituted the core comparison between the experimental group (interacting with the \textit{ImmuniFraug} simulation) and the control group (reading a static, text-based article). 2) \textbf{Fraud Scenario}, which included ten types to assess the generalizability of our findings across various contexts. Participants were automatically and randomly assigned to one of the 20 resulting conditions (10 fraud types × 2 modalities), ensuring each individual was exposed to only one scenario and one condition. The example reading material for control group is in Appendix~\ref{controlgroup} 

The experimental procedure unfolded as follows. First, participants in the experimental group engaged with the \textit{ImmuniFraug} system, which presents realistic cyber fraud scenarios through voice-driven conversation, AI-generated visuals, and dynamic narrative branching. In parallel, participants in the control group were instructed to read an official fraud awareness article and a real-case summary corresponding to the same fraud type. This control material was compiled and published by the Ministry of Public Security of China \cite{MPS2024stats}, ensuring that both groups received equivalent thematic content, differing only in the modality of presentation (interactive simulation vs. static text).

Following the intervention, all participants completed a standardized fraud awareness test to assess their recognition and understanding of fraud mechanisms. This test consisted of eight multiple-choice questions, with its structure informed by established methodologies \cite{chung2023reducing, IFEC2021}. Subsequently, participants in the experimental group only were directed to complete two additional sets of questionnaires. The first was a modified version of the Narrative Transportation Scale \cite{green2000role} to measure their psychological immersion. The second was a custom-designed questionnaire assessing four constructs on a 5-point Likert scale: Perceived Fraud Awareness, Perceived Self-Efficacy, Behavioral Intention, and a Comparative Evaluation against traditional methods. The full custom questionnaire can refer Appendix~\ref{CustomQuestionnaire}

To ensure sufficient statistical power for detecting effects within the experimental group's self-report measures, we conducted an a priori power analysis using G*Power. The analysis indicated that a minimum of 290 participants in the experimental group was required to detect a small-to-medium effect size ($d=0.3$) with 95\% power.

\begin{table}[ht]
\centering\small

\caption{Demographic information of participants}
\label{tab:demographics}
\begin{tabularx}{\linewidth}{>{\RaggedRight\arraybackslash}Xll}
\toprule
\textbf{Category} & \textbf{Subcategory} & \textbf{Responses} \\
\midrule
\multirow{3}{*}{Gender} 
 & Male & 296 (34.9\%) \\
 & Female & 550 (64.9\%) \\
 & Non-binary & 1 (0.1\%) \\
\midrule
\multirow{2}{*}{Group}
 & Control & 455 (53.8\%) \\
 & Experiment & 391 (46.2\%) \\
\midrule
\multirow{2}{*}{Age} 
 & 18--20 & 358 (42.6\%) \\
 & 21--25 & 482 (57.4\%) \\
\midrule
\multirow{2}{*}{Anti-fraud education} 
 & Yes & 778 (92.0\%) \\
 & No & 68 (8.0\%) \\
\midrule
\multirow{2}{*}{Experienced fraud} 
 & Yes & 349 (41.3\%) \\
 & No & 497 (58.7\%) \\
\midrule
\multirow{2}{*}{Someone nearby was defrauded}
 & Yes & 534 (63.1\%) \\
 & No & 312 (36.9\%) \\
\bottomrule
\end{tabularx}
\end{table}

To gain deeper insight into participants' subjective experiences, we invited all study respondents to participate in optional follow-up interviews. From those who expressed interest and test results, we selected a diverse sample of 15 individuals for semi-structured interviews. To ensure that the qualitative insights from this subsample were not biased by varying levels of immersion across different scam types, we conducted a preliminary statistical analysis on their self-reported immersion scores. A between-group comparison revealed no statistically significant differences in narrative engagement ($p = 1.00$). This result confirms that the immersive experience was consistent across the participants selected for interviews, supporting the validity of treating their qualitative feedback as a cohesive dataset. 
We conducted the interview through Zoom, were designed to explore participants' perceptions of fraud, their engagement with the system, and its overall usability, with a particular focus on the elements of \textit{ImmuniFraug} that contributed to narrative immersion. All sessions were audio- and video-recorded and subsequently transcribed verbatim. The qualitative data were then analyzed using thematic analysis. Two researchers independently performed open coding to identify emergent themes from the data. Discrepancies in coding were resolved through discussion to ensure inter-rater reliability. Finally, all raw data, codes, and thematic structures were translated into English to facilitate reporting and broader analysis. The example interview protocol can refer Appendix~\ref{ExampleInterviewProtocol}



\subsection{Participant Recruitment}
We recruited participants through a comprehensive online campaign leveraging social media platforms and internal university communication channels. The recruitment announcement provided a detailed overview of the research, the estimated time commitment, and participant incentives. To verify student status and mitigate the impact of automated bot responses, participants were required to manually provide their academic majors; this information was utilized solely for verification purposes and was not stored for subgroup analysis. The gender distribution of our sample, characterized by a higher proportion of female participants, accurately reflects official public demographics of the target undergraduate population in the region of study\footnote{https://m.bjnews.com.cn/detail/1756191834129050.html}.

A total of 1,095 individuals accessed the study link hosted on the online survey platform. Of these, 846 participants completed the full procedure and were included in the final analysis, representing an attrition rate of 22.7\%. The survey platform automatically and randomly assigned participants to either the control ($n=455$) or the experimental ($n=391$) condition. The slight variance in final group sizes is attributed to minimal participant attrition, which remained comparable between the experimental and control groups. This suggests that our ethical withdrawal mechanism—allowing participants to exit at any stage without consequence—was effective and did not introduce systematic bias. As compensation, all participants received a monetary reward or an equivalent-value coupon. Detailed participant demographics are presented in Table \ref{tab:demographics}.

\section{Results}

\subsection{RQ1: Effectiveness of LLM-based Simulation in Improving Cyber Fraud Awareness}

\subsubsection{Preliminary Between-Group Analysis}
To evaluate whether the LLM-based fraud scenario simulation effectively enhanced fraud awareness among undergraduate students, we conducted a series of independent samples t-tests. Participants were divided into two groups: the experimental group, which engaged in an immersive fraud simulation powered by a large language model, and the control group, which did not undergo any such intervention. The independent samples t-test was deemed appropriate for this context, as it is specifically designed to compare the means of two independent groups to determine whether there is a statistically significant difference between them\cite{field2024discovering}. Given that our primary objective was to assess the differential impact of the LLM intervention on fraud awareness scores, this method provided a robust framework for hypothesis testing.
 
To ensure the validity of the t-test results, we examined the three key assumptions underlying its application: normality, homogeneity of variance, and independence of observations. The detailed results can be found in Table \ref{tab:fraud_ttest_results}. First, normality was assessed and confirmed for all items, as indicated by the "Y" markers in the "Normality" column of the table. This suggests that the data for each test item in both groups approximated a normal distribution, satisfying the requirement for parametric testing. Second, Levene’s Test for Equality of Variances was conducted for each item to test the assumption of homogeneity of variances\cite{levene1960robust}. The p-values of Levene’s test were greater than 0.05 for most items, indicating that the variances between groups were generally equal. Notable exceptions include ``Impersonation of Acquain-
tance'' (F = 6.882, $p = 0.01$), ``Fake In-Game Trader'' (F = 6.689, $p = 0.011$), and the overall score (F = 6.888, $p = 0.009$), where the assumption of equal variances was violated. Moderate violations of the equal variance assumption in an independent samples t-test are generally acceptable because the test is considered robust, particularly when the sample sizes are equal or nearly equal. In such cases, the impact on the Type I error rate is minimal, and the validity of the test remains largely unaffected. This robustness has been supported in prior research\cite{ruxton2006unequal,zimmerman2004note}, suggesting that strict adherence to the homogeneity of variance assumption is not always necessary unless sample sizes are highly unequal. The third assumption, independence of observations, cannot be tested statistically. However, it was rigorously maintained in the experimental design, as participants were randomly assigned to groups and completed the assessments independently.

It is worth noting that fraud prevention education in China is currently highly pervasive \cite{CAC2025}, with 92\% of our participants reporting prior exposure to anti-fraud training. As a result, participants entered the study with relatively high baseline awareness, which may have constrained the observable effect size and limited the potential for highly significant between-group differences. 

The results of the independent samples t-tests are presented in the table. Most item-level comparisons between the experimental and control groups yielded non-significant results ($p > 0.05$), indicating no substantial differences in fraud awareness across individual fraud scenarios. However, the total fraud awareness score demonstrated a statistically significant difference between the two groups (t = 2.229, df = 803.194, $p = 0.026$), with the experimental group showing a higher mean score (M = 35.06, SD = 4.59) compared to the control group (M = 34.73, SD = 6.76). This suggests that, while specific item-level effects may be limited, the overall intervention had a small but statistically significant positive impact on general fraud awareness. The effect size for the total score, as measured by Cohen’s d = 0.151, falls within the small range, indicating a modest practical effect\cite{cohen2013statistical}. Furthermore, we conducted additional t-tests within the experimental group to examine differences across simulated scam categories using our custom fraud awareness scale. These analyzes did not reveal significant differences among the categories, indicating a relatively uniform perception of fraud awareness regardless of the type of scam.

\begin{figure}[htbp]
    \includegraphics[width=\columnwidth]{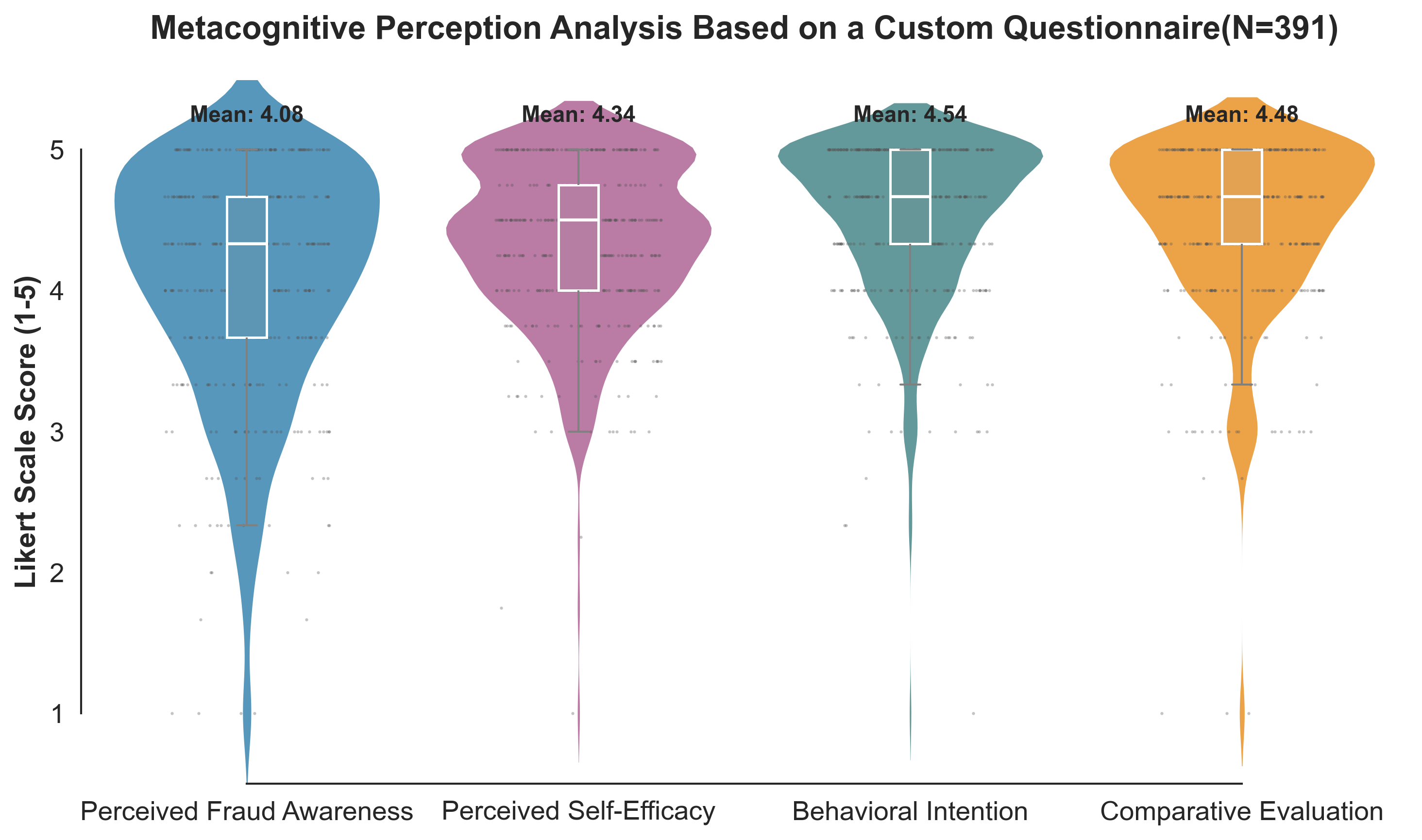}
    \caption{Distribution of subjective scores across the four key constructs. The violin plots illustrate the kernel density estimation of the data. Inside, the box plots represent the median (thick line) and interquartile range (IQR), while the overlaid gray jittered points depict the individual perceptual ratings of the 391 participants. Refer to [\ref{CustomQuestionnaire}] for the detailed questionnaire items.}
    \label{fig:custdata}
\end{figure}

Custom questionnaire feedback from the experimental group ($N=391$)  further substantiates the intervention's impact on participants’ internal psychological states. As illustrated in the Metacognitive Perception Analysis (Figure \ref{fig:custdata}), scores across all evaluated constructs were robustly high, reflecting a strong positive shift in subjective preparedness. Specifically, participants reported high Perceived Fraud Awareness ($M = 4.08$) and Perceived Self-Efficacy in their ability to detect deceptive tactics ($M = 4.34$). Most notably, the intervention elicited exceptionally strong Behavioral Intentions to adopt protective measures ($M = 4.54$) and a favorable Comparative Evaluation of the simulation’s effectiveness relative to traditional methods ($M = 4.48$). With an aggregate mean score of 4.36 across all dimensions, these results indicate that ImmuniFraug successfully fosters the metacognitive engagement and psychological resilience required to resist sophisticated social engineering. While item-level statistical variances exist within the broader study, this subjective data reinforces our conclusion that the interactive simulation is a highly effective tool for anti-fraud education. Additional statistical breakdowns are accessible in Supplementary Material. 

\begin{table}[ht]
\centering
\caption{LMEM Results for Fraud Awareness Scores. The model utilizes Restricted Maximum Likelihood estimation. C(Gender) [Female] represents the score difference of female participants relative to males(One non-binary participant was included in the analysis but not listed separately due to the minimal sample size). Age Group (21--25) indicates the difference of the 21--25 group relative to the 18--20 group. \textbf{Random Effects} account for the baseline intercept variance across the ten distinct fraud scenarios.}
\label{tab:lmem_results}
\small
\begin{tabular}{l r r r r}
\toprule
\textbf{Fixed Effects} & \textbf{Coef.} & \textbf{Std.Err.} & \textbf{z} & \textbf{P $> |$z$|$} \\
\midrule
Intercept (Baseline) & 35.545 & 1.035 & 34.337 & 0.000 \\
Condition (Intervention) & \textbf{0.859} & \textbf{0.385} & \textbf{2.228} & \textbf{0.026} \\
C(Gender) [Female] & 0.547 & 0.402 & 1.360 & 0.174 \\
Age Group (21--25) & -0.445 & 0.404 & -1.101 & 0.271 \\
Educational Exposure & -0.709 & 0.712 & -0.996 & 0.319 \\
Personal Experience & -0.597 & 0.428 & -1.396 & 0.163 \\
Social Observation & 0.051 & 0.435 & 0.116 & 0.907 \\
\midrule
\textbf{Random Effects} & \textbf{Var.} & \textbf{Std.Dev.} & & \\
\midrule
Scenario (Group Var) & 4.233 & 2.057 & & \\
\bottomrule
\end{tabular}
\end{table}

\subsubsection{Effectiveness Evaluation via Linear Mixed-Effects Modeling}

To rigorously evaluate the intervention effect of \textit{ImmuniFraug} while addressing the heterogeneity among the ten distinct fraud scenarios, we transitioned from traditional comparative statistics to a more robust \textbf{LMEM}\cite{bates2015fitting}. This approach allows us to isolate the intervention's net impact by statistically partitioning the variance into fixed and random components.

The dependent variable is the participant's total awareness score. The primary fixed effect is the \textbf{Experimental Condition} (0 = Control, 1 = Experimental). To account for individual background differences and potential "ceiling effects" in baseline knowledge, we incorporated five fixed-effect covariates:
\begin{itemize}[noitemsep, topsep=2pt]
    \item \textbf{Demographics}: Gender (coded as a categorical variable) and Age Group (binary: 18–20 vs. 21–25).
    \item \textbf{Prior Fraud Exposure}: Three specific binary indicators reflecting the participants’ baseline preparedness, which were collected as part of the basic information assessment prior to their participation in the experiment: (1) Previous anti-fraud education; (2) Personal history of fraud victimization; and (3) Fraud experiences within social circles.
\end{itemize}

To address the reviewer’s concern regarding scenario-specific variability, we treated the \textbf{Scenario Type} as a random effect (random intercepts). The formal model is specified as:

\begin{equation}
\resizebox{0.9\columnwidth}{!}{%
    $Score_{ij} = \beta_0 + \beta_1 Condition_i + \beta_2 Gender_i + \beta_3 Age_i + \sum_{k=1}^3 \gamma_k PriorExp_{ki} + u_j + \epsilon_{ij}$%
}
\end{equation}

where $\beta_1$ represents the intervention's fixed effect, $u_j \sim N(0, \sigma_u^2)$ accounts for the random intercept of scenario $j$, and $\epsilon_{ij}$ is the residual error.

The LMEM analysis was conducted on $N=846$ valid observations across ten scenarios (mean group size = 84.6). As summarized in Table \ref{tab:lmem_results}, the model achieved convergence with a Log-Likelihood of -2651.12. The analysis revealed a statistically significant fixed effect for the \textbf{Experimental Condition} ($\beta = 0.859, SE = 0.385, z = 2.228, p = 0.026$). This result indicates that, after controlling for demographic factors and all dimensions of prior fraud exposure, the \textit{ImmuniFraug} group significantly outperformed the control group. The positive coefficient confirms the system's effectiveness in enhancing fraud awareness beyond static educational materials. Crucially, the intervention's efficacy remained robust across different participant backgrounds. Demographic covariates: Gender ($p = 0.174$) and Age Group ($p = 0.271$), did not significantly influence the scores. Similarly, prior experiences, including educational background ($p = 0.319$), personal victimization ($p = 0.163$), and social observation ($p = 0.907$), showed no statistically significant predictive power for the final outcome. The significant \textbf{Group Variance} ($\sigma_u^2 = 4.233$) confirms the presence of baseline differences across fraud types, validating our decision to utilize a mixed-effects framework to isolate the scenario-independent effect of the intervention.



\subsection{RQ2: Perceptions of the Simulation and its Metacognitive Impacts}


To address RQ2, we examined participants' perceptions of the simulation and its metacognitive impacts. We first measured narrative immersion using an adapted version of the Transportation Scale developed by Green et al.~\cite{green2000role}, with scores ranging from 7 to 77. The results showed that the experimental group achieved a mean transportation score of 56.95, indicating a relatively high level of narrative immersion. This suggests that the \textit{ImmuniFraug} was effective in engaging undergraduates and fostering psychological transportation into the simulated scam scenarios.

To understand the mechanisms driving this strong immersion results and to explore its subsequent effects on learning, we conducted a thematic analysis of semi-structured interviews. Table \ref{tab:participant_summary} provides detailed information about the interview participants. The codebook is provided in Table \ref{tab:codebook_final}. The analysis revealed a psychological process that unfolds across three core themes, which structure the remainder of this section: (1) Perceived Realism and Immersion, which explains why the framework experience was engaging; (2) Cognitive and Emotional Engagement, which details how participants were involved; and (3) Learning Outcomes and Metacognitive Reflection, which describes what they ultimately gained, directly explains how this experience shaped their understanding.

\begin{table}[htbp]
\centering
\scriptsize 
\caption{Participant Background and Fraud Experience Summary. RWSE: Real-World Scam Experience; SiS: Scammed in Simulation.}
\label{tab:participant_summary}

\begin{tblr}{
    width = \columnwidth,
    colspec = { 
        Q[c,m, 0.3cm]  
        Q[c,m]         
        X[0.8,c,m]     
        X[1.3,c,m]     
        X[1.9,c,m]     
        Q[c,m]         
        Q[c,m]         
    },
    row{1} = {font=\bfseries},
    hlines = {0.5pt, gray8}, 
    colsep = 2pt,  
    rowsep = 0.5pt, 
}
ID & Gen. & Grade & Major & Fraud Type (EFT) & RWSE & SiS \\ 
P1 & F & Junior & Computer Science & Fake Logistics Agent & Peer & No \\ 
P2 & M & Junior & Computer Science & Fake Police/Gov. Officer & Self & No \\ 
P3 & F & Soph. & Electronic Eng. & Impersonation of Acquaintance & Self & No \\ 
P4 & F & Senior & Law & Online Romance Scam & Self & No \\ 
P5 & F & Fresh. & HCI & Impersonation of Acquaintance & Self & Yes \\ 
P6 & F & Soph. & Early Child. Edu. & Fake Credit Report Officer & Encount. & Yes \\ 
P7 & F & Soph. & Software Eng. & Online Romance Scam & Encount. & No \\ 
P8 & M & Fresh. & Info. Systems & Fake Loan Officer & Self & No \\ 
P9 & F & Soph. & Agronomy & Fake Loan Officer & Self & No \\ 
P10 & F & Senior & Journalism & Fake Police/Gov. Officer & Encount. & No \\ 
P11 & F & Junior & Nursing & Impersonation of Acquaintance & Encount. & No \\ 
P12 & M & Senior & Finance & Fake Investment & Peer & No \\ 
P13 & M & Fresh. & Design & Fake Task Rebates & Self & No \\ 
P14 & M & Senior & Computer Science & Fake In-Game Trader & Peer & No \\ 
P15 & F & Senior & Marketing & Fake Credit Report Officer & Peer & No \\ 
\end{tblr}
\end{table}

\subsubsection{Perceiving the Interactive Experience}

A primary finding was \textit{ImmuniFraug}'s ability to induce a strong sense of narrative immersion. Participants reported feeling as though the system had `come alive,' fostering a genuine desire to continue the interaction (P1). This powerful sense of immersion was underpinned by three key factors:

\paragraph{Perceiving the realism forged through scenario and tone.}
Participants generally perceived the scam scenarios and processes simulated by \textit{ImmuniFraug} as highly realistic, especially those who had personally encountered fraud in the past. For instance, P10 remarked:\textit{``I’ve been scammed before, and I felt that this experience was very similar to what I went through,it really felt quite real.'' } 
In addition to the scenario design, participants noted that the tone of interaction further enhanced the sense of realism. The system’s communication style was described as convincingly aligned with real scam encounters, not overly polite or robotic, but intentionally curt or pushy. As P1 pointed out: \textit{``The tone wasn’t overly friendly. It actually reminded me of an impatient customer service rep it felt very authentic.'' }

\paragraph{Experiencing the calculated persuasion through progressive tactics and emotional exploitation.}
Participants noted that the \textit{ImmuniFraug} employed highly effective and calculated persuasive strategies designed to gradually ensnare them in simulated fraud scenarios. Rather than presenting abrupt or implausible demands, the system adopted a step-by-step approach, fostering a gradual narrative buildup that progressively increased compliance over time. This design mirrored real-world social engineering tactics that create high-pressure decision environments, often by invoking scarcity or urgency, thereby disrupting calm deliberation. The system's ability to subtly guide user behavior incrementally aligned with prior research on large language models~\cite{10.1145/3531146.3533088}. \textit{ImmuniFraug} further enhanced this realism by adopting the first-person perspective of the fraudster, authenticating the full course of targeted scam types. As P4 articulated, \textit{``It didn’t make any sudden or harsh demands. Instead, it led you in slowly like starting with a request that’s clearly unreasonable, then gradually lowering the bar until you feel compelled to agree.'' }(P4)

Furthermore, participants recognized how \textit{ImmuniFraug} strategically exploited their emotional vulnerabilities. The system tailored its responses to trigger anxiety around potential legal or financial consequences, effectively pushing undergraduates toward compliance. This included tactics of emotional exploitation, such as attempting to isolate users from external support and systematically eroding their confidence in their own memory and judgment through insistent messaging. The simulation also leveraged need exploitation, tapping into undergraduates' desires for recognition, career advancement, or social acceptance to lower their defenses, making them more susceptible to manipulation when emotional incentives aligned with personal aspirations. P8's experienced: \textit{``It really sounded like it was trying to help. It warned me that if I contacted the platform myself and they didn’t respond quickly, I might lose time, or worse if it affected my credit score, I’d regret it deeply. That kind of consequence-driven talk is exactly what real scammers use.''} These adaptive and dynamic deceptive techniques were key to reproducing the psychological pressure and nuanced persuasion characteristic of actual scam interactions, contributing significantly to the perceived realism of the simulation.

\paragraph{Enhancing persuasion and memory retention through concrete, contextual scenarios.} 
Many participants contrasted the \textit{ImmuniFraug} experience with traditional anti-fraud outreach formats, such as lectures or informational videos. They consistently described those conventional approaches as passive, slogan-based, and largely ineffective, primarily due to their lack of personal involvement. This aligns with prior research indicating that traditional ``tell-only'' strategies, reliant on static content, often fail to cultivate the critical thinking and behavioral resilience needed for real-world threats. In stark contrast, participants perceived the simulated interaction in \textit{ImmuniFraug} as a deeply participatory experience that actively prompted reflection and enhanced memory retention. This immersive, conversation-driven design elicits a psychological state of ``transportation,'' involving focused attention, vivid mental imagery, and emotional engagement. Such engagement allows users to process messages with fewer counter-arguing interruptions and encode salient cues more deeply, making the recognition of "red flags" more automatic. As P7 remarked: \textit{``When it’s a lecture, most people don’t really listen, they’re doing their own thing. But this one felt like I was personally involved. It stayed with me because it was like making my own decisions and experiencing the consequences. That’s what made me reflect.''}

\paragraph{Leveraging multimodal cues to induce temporal pressure and narrative immersion.}

To induce a compelling sense of both narrative immersion and temporal pressure, the \textit{ImmuniFraug}  leveraged auditory and visual cues, each fulfilling a distinct function. The auditory modality's contribution was twofold. First, it was instrumental in establishing a deep narrative immersion by creating a rich, atmospheric context. As P6 highlighted, the experience felt akin to a game:  \textit{``The voice definitely made it more immersive. I think voice is essential for creating the atmosphere. If there were even a bit of narration, it would feel like I was playing a mystery role-playing game.''} Second, beyond atmosphere, the voice channel introduced a critical element of temporal pressure, simulating the urgency of a real-world scam by compelling immediate responses. P12's feedback underscores this point: \textit{``Voice makes it better. It doesn’t give you much time to think. It forces you to respond right away. That’s what makes it more immersive.''}.

While the auditory channel drove the simulation's pacing and atmosphere, the visual modality was essential for grounding the experience in authenticity. The system employed intentionally crafted avatars tailored to the social context of each fraud type, a design choice that proved highly effective at triggering real-world associations. As P15 described, this visual tailoring lent significant credibility to the interaction: \textit{``The character made me think of someone selling houses, like those bank employees pushing insurance. It just gave me that vibe very convincing.''} This association demonstrates how visual cues fortified the simulation's realism, making the characters and their intentions feel more genuine. Ultimately, it was this deliberate combination of auditory cues driving pressure and immersion, and visual cues reinforcing authenticity, that created a holistically effective simulation.

\paragraph{Crafting adaptive deception to enhance realism and immersion.}
Participants emphasized that, unlike traditional scripted scenarios, \textit{ImmuniFraug} was able to dynamically adapt its responses in real time based on undergraduates’ input. This responsiveness allowed the system to craft increasingly personalized and targeted replies, which significantly enhanced the sense of realism and user immersion. P3 commented: \textit{``It adjusted its responses based on what I said, generating replies that felt more emotionally pressuring or manipulative. That made it easier to get drawn in. It really heightened the sense of immersion.''} Such adaptive deception made the interaction feel less predictable and more convincing, leaving participants with a deeper and more memorable experience compared to static, pre-defined scripts.

\subsubsection{Fostering Deep Cognitive and Emotional Engagement}

\paragraph{Compelling sustained attention by demanding cognitive engagement.}
Participants reported that the interactive nature of \textit{ImmuniFraug} enhanced their attentional focus, requiring them to remain mentally engaged throughout the experience. Unlike traditional materials such as posters or videos, which were often described as passive and easily dismissed, the system’s requirement for active input and response kept undergraduates cognitively involved. As P1 explained, \textit{``It really doesn’t let you drift off. It makes you stay focused. When I see a flyer or a video, I glance at it and feel like I got the point. But here, I had to keep responding, keep interacting, and that actually helped me concentrate and really try to understand what was going on.''}

\paragraph{Fostering curiosity for exploratory learning.}
Beyond focused attention, many participants described the experience as novel and exploratory, sparking a strong sense of curiosity. The interactive format encouraged undergraduates not only to follow the expected script, but to intentionally test boundaries, wondering what might happen if they responded differently, even if they knew it was simulated. This reflects a form of safe experimentation, where participants were willing to ``push'' the interaction precisely because it was free of
 consequence. As P5 shared, \textit{``I was curious, if I responded differently, what would the scammer say? I knew there were boundaries, but I still wanted to test them a little. In real life, I’d just hang up the phone. But with this chatbot, I wanted to try different possibilities.''} By providing a psychologically safe environment, the system enabled `playful risk-taking', making the experience intrinsically motivating for undergraduates and driven by curiosity rather than compliance. This hands-on engagement allowed participants to gain a deeper, first-hand understanding of scammer strategies instead of relying on passive warnings.

\paragraph{Simulating time pressure to impair rational judgment.}
One of the key strengths of \textit{ImmuniFraug} lies in its ability to simulate the time-sensitive pressure often present in real-world scam encounters. Unlike traditional learning tools, such as multiple choice quizzes that allow for deliberate reasoning. The interactive format demands immediate user responses, reducing the opportunity for careful reflection. This design more accurately mirrors the high-stress, fast-paced nature of actual scam situations. Several participants noted that under time constraints, their ability to assess and respond rationally was significantly impaired. P2 reflected: \textit{``If it were just a simple multiple-choice question, I’d instantly know which option is correct. But in a scenario like this, it’s hard to make the right decision quickly.''}

\paragraph{Replicating emotional manipulation via isolation and doubt.}
Participants reported that the emotional manipulation tactics used in \textit{ImmuniFraug} closely mirrored those of real-life scams, particularly in the way they sought to exert psychological control over the target. The simulation effectively recreated the emotional pressure and disorientation often used by scammers to isolate and destabilize their victims. One common tactic identified was the deliberate effort to isolate undergraduates from external sources of support or validation. By discouraging communication with others and monopolizing the information flow, the system recreated a sense of psychological enclosure. As P3 described: \textit{``The scam tried to isolate you. To cut you off from the outside world and trap you in their logic. It’s just you and the scammer, and all your information comes from them.'' } 

In addition to isolation, the simulated scam also worked to erode undergraduates' confidence in their own memory and judgment. Through repeated and insistent messaging, participants began to question their actions and whether they might have unknowingly triggered the situation themselves. P15 reflected: \textit{``They kept insisting on this issue, and after a while, I started to doubt myself. Like, maybe I did click something by accident, such as some apps activate services without telling you. The repetition really made me question whether I had done something wrong.''}. These emotionally manipulative strategies demonstrated the simulation’s ability to evoke realistic psychological responses, providing participants with a deeper appreciation of the subtle yet powerful tactics used in real-world fraud.

\paragraph{Demonstrating vulnerability by exploiting latent needs.}
Participants expressed the \textit{ImmuniFraug} effectively demonstrated how scams exploit the specific psychological and situational needs of their targets, such as the desire for recognition, career advancement, or social acceptance, to bypass critical thinking. The system mimicked this by presenting scenarios tailored to participants' latent motivations, thereby making them more susceptible to manipulation. When a scam narrative aligned with personal ambitions, maintaining critical distance became significantly more difficult. P5 articulated \textit{``You might really want to gain your supervisor’s trust, or you’re eager to get promoted that kind of desire makes you drop your guard.''} 
By exploiting such needs, the simulation highlighted a key vulnerability in scam susceptibility: when emotional incentives align with personal aspirations, undergraduates may unconsciously prioritize opportunity over caution. This insight helped participants better recognize the subtle influence of motivational triggers in real-world fraud contexts.

\paragraph{Building empathy by shifting perspective.}
The simulation deepened empathy by shifting participants from a detached, third-person observer to the first-person position of a target. Many acknowledged holding prior biases that scam victims were simply naive. Through \textit{ImmuniFraug}, they gained a more nuanced understanding of how emotional pressure can impair anyone's judgment. P9 reflected on this perspective shift: \textit{``I think people usually view scam victims from a detached, third-person standpoint. But when it happens to you, when you get that call or message, and you suddenly realize your personal info is exposed or money is lost. It’s genuinely stressful and overwhelming.''}

\paragraph{Triggering cognitive dissonance to challenge overconfidence.}
Participants reported the simulation consistently triggered cognitive dissonance \cite{festinger1962cognitive}, the mental discomfort experienced when participants' belief in their own invulnerability clashed with their demonstrated susceptibility during the exercise. Participants experienced cognitive dissonance, which a conflict between their prior confidence and their felt vulnerability during the simulation. Most entered the experience believing they would easily resist, only to find themselves momentarily caught off guard or manipulated. This contradiction forced a re-evaluation of their own assumptions about their preparedness. P4 candidly reflected on this state of cognitive entrapment: \textit{``When your thinking is trapped in that situation, it's hard to step outside and see clearly.''} 

\subsubsection{Enhancing Learning Outcomes and Metacognitive Reflection}

\paragraph{Internalizing fraud tactics through simulated engagement}
The \textit{ImmuniFraug} experience deepened participants' cognitive understanding of fraud tactics. Through active engagement, they became more attuned to the subtle cues, patterns, and psychological mechanisms of real-world fraud. Repeated exposure helped internalize key red flags, as P1 noted: \textit{``The cases were very vivid. After going through them a few times, I started to recognize repeated scam scripts. It gave me that ‘ah, so that’s how it works’ kind of realization.''} Knowing the interaction was a simulation encouraged participants to actively scrutinize the scam's logic, deepening their awareness through metacognitive monitoring~\cite{rivers2020measuring}. They deliberately looked for inconsistencies, such as procedural flaws that contradicted legal or normative expectations. As P14 observed: \textit{``They said there was a dedicated team bypassing the official platform to conduct direct transactions, that’s clearly non-compliant no matter how you frame it.''} 

Others learned to recognize behavioral cues like manufactured urgency. P9 shared that the scammer's tone and exaggerated consequences were telling: \textit{``I think tone is a really critical indicator. Throughout the conversation, the simulated scammer sounded extremely anxious. Like telling me not to hang up and exaggerated the consequences, saying it could affect my civil service exams or even future childbirth. That level of dramatization is a red flag.'' } Collectively, these insights suggest that \textit{ImmuniFraug} not only educates undergraduates about scam content, but also fosters critical thinking and situational awareness that may transfer to real-world encounters.

\paragraph{Activating metacognitive reflection and recall.}
The \textit{ImmuniFraug} simulation not only taught undergraduates how scams operate, but also encouraged deeper personal reflection and memory recall, two cognitive processes crucial for long-term learning and behavioral change. Participants were prompted to evaluate their own vulnerabilities and reconsider how they might respond in real situations. The immersive nature of the simulation highlighted individual blind spots, helping undergraduates recognize the psychological triggers that might make them more susceptible to scams. As P5 reflected:\textit{``When it comes to scams, we really need to develop better recognition skills. You shouldn’t risk losing a lot just because of the temptation of small, easy gains.''} In several cases, the scenarios triggered vivid recollections of actual scams participants had encountered, reinforcing the perceived realism and relevance of the experience. For example, P10 shared:\textit{``When I got scammed before, it felt really similar to this. They also asked me to provide certain information, almost identical to the options the AI gave me.''} 
By linking abstract learning to personal context, the simulation activated key metacognitive processes~\cite{flavell1979metacognition,schraw1995metacognitive}. This self-referential thinking, prompted by reflection and memory recall, reinforced immediate awareness and supported future behavioral adaptation.

\paragraph{Eliciting awareness of AI-driven fraud.}
Through the \textit{ImmuniFraug} experience, participants gained a heightened awareness of the capabilities of LLMs and, more importantly, the potential risks they pose when misused for malicious purposes. The realism and interactivity of the simulation led undergraduates to reflect on how advanced AI systems could be exploited to conduct more convincing and scalable scams. There is a notable convergence between the participants’ concerns and the priorities highlighted in contemporary AI safety research\cite{li2025security}.

Participants expressed growing concerns about the increasing sophistication of AI-generated deception, particularly when combined with technologies like deepfakes and voice synthesis. As P7 warned:\textit{``AI feels genuinely deceptive now. It can even fake your appearance, like during a video call, it could replicate your face. That makes it really dangerous. It could impersonate a family member and ask you for money. It’s basically impossible to guard against that.''}

\subsection{RQ3: Limitations and Requirements}
While participants found \textit{ImmuniFraug} highly effective, they also identified key limitations and requirements to improve its authenticity and impact of the intervention.

\paragraph{Improving vocal realism to enhance immersion.}
Participants noted the system's synthetic voice quality as an area for improvement, as its artificial nature could diminish the immersive effect. The delivery lacked the natural prosody and emotional nuance of human speech, making it sound mechanical. As P2 remarked: \textit{``The voice still felt mechanical. It opened with something like `Finally, you picked up', saying it's from some insurance or loan company but the tone didn’t feel authentic. It needs to sound more real.''} 

\paragraph{Extending conversational memory for longitudinal scams.}
A structural limitation identified was the model's inability to sustain long-term conversations due to token limits. This restricts simulations to short, self-contained interactions, preventing the replication of prolonged manipulation strategies (e.g., romance or investment scams) that unfold over extended periods. P4 illustrated this challenge: \textit{``In real life, if someone’s trying to manipulate a romantic partner into investing, they build it up much more gradually. But since this is an experiment, you obviously can’t simulate a three-day conversation.''} 

\paragraph{Integrating multimedia for multimodal deception.}
Some participants highlighted the absence of visual elements as a significant gap in realism. Relying solely on text or voice fails to replicate the multi-modal nature of modern scams, which often leverage fake websites, doctored images, or phishing links to appear credible. As P6 explained: \textit{``A lot of the time it was just text. Without images or anything like that, it didn’t feel as real as a scam where they’d send you a fake screenshot or a phishing link.''}

\paragraph{Targeting vulnerable groups while stressing universal relevance.} 
Participants recommended targeting individuals with limited social and digital experience, such as younger students and older adults, who could benefit most from building awareness in a safe environment. However, they also stressed the universal nature of susceptibility, arguing that the simulation is valuable for everyone regardless of age or education. P11 powerfully concluded that the tool helps identify personal vulnerabilities before they can be exploited: ``I think everyone should try it. Just because you haven’t been scammed doesn’t mean you’re immune. It might just be that the right scam hasn’t come along yet. This kind of simulation helps you find out which type of scam you’re most likely to fall for, so you can avoid it in the future.''

\section{Discussion}

\subsection{Interpreting Impact: Immersion, Awareness, and Metacognition}
Our findings suggest that \textit{ImmuniFraug} works not merely by adding more anti‑fraud facts to undergraduates' memory, but by reconfiguring how undergraduates appraise, feel, and decide under social pressure. The pattern in RQ1-RQ3: consistent score advantages despite high prior training, robust narrative engagement in scams, and qualitative reports of attentional focus, urgency, and reflection points to a mechanism that couples immersive experience with metacognitive consolidation rather than traditional content-centric instruction.

\textbf{Narrative immersion as a learning driver.} The immersive, conversation‑driven scenarios appear to elicit the psychological state of transportation, focused attention, vivid mental imagery, and emotional engagement, long linked to changes in beliefs and evaluations in narrative contexts~\cite{green2000role,van2014extended}. When undergraduates ``enter'' the story, they process messages with fewer counter‑arguing interruptions and encode salient cues more deeply~\cite{slater2002entertainment,appel2010transportation}, which can make later recognition of red flags more automatic. This theorized pathway aligns with participants’ descriptions of being ``drawn in,'' feeling time pressure, and experiencing the interaction as a live negotiation rather than a lesson. Prior work shows that such transportation increases story‑consistent beliefs and evaluations, helping explain why experiential simulations can outperform passive warnings in shaping judgments under uncertainty~\cite{moyer2008toward,green2000role}.

\textbf{Affective arousal timed to a teachable moment.} Many participants emphasized how the simulated urgency and social pressure made it harder to deliberate calmly mirroring the affective dynamics of real scams. Rather than treating emotional arousal as a confound, \textit{ImmuniFraug} concludes each scenario with a structured debrief. Although the emotional trajectory may vary across sessions, ending with reflective guidance allows the system to leverage residual affect-aligning with the notion of the teachable moment, wherein salient experiences can increase receptivity to risk information when followed by timely reflection. This sequencing resonates with the teachable moment, which argues that salient, affect‑laden events can enhance receptivity to risk information and motivate behavior change when reflective guidance is offered immediately afterward~\cite{lawson2009teachable}. Memory research on the peak–end rule further suggests that what undergraduates remember (and thus reuse as a heuristic) is disproportionately shaped by the peak intensity and the way the episode ends, underscoring the value of closing the scenario with structured reflection that names the manipulation tactics just experienced~\cite{kahneman1993more}.

\textbf{From awareness to action via metacognition and self‑efficacy.} The debrief stage is intentionally metacognitive. It asks undergraduates to monitor how their attention shifted, label the persuasive tactics, and plan if then responses for future encounters. Across learning sciences, metacognitive regulation is associated with stronger transfer and performance, in part because it calibrates self‑efficacy people's beliefs in their capability to execute the behaviors required under pressure~\cite{bandura1977self}. As efficacy increases through guided mastery and explicit strategy rehearsal, individuals are more likely to initiate protective actions and sustain them when confronted by aversive cues (e.g., threats, time scarcity)~\cite{halmo2024metacognition}. This provides a plausible bridge between our observed attitudinal gains and the behaviors undergraduates report intending to adopt. 

\textbf{Inoculating against manipulation through experiential reframing.}  
Rather than relying on didactic warnings or passive informational campaigns, \textit{ImmuniFraug} engages undergraduates in a simulated, high-stakes interaction that mirrors real-world fraud scenarios. This immersive experience is followed by guided reflection that prompts undergraduates to identify manipulation tactics and mentally rehearse protective responses. By pairing affective engagement with reflective reframing, the system helps consolidate durable anti-fraud heuristics and enhances undergraduates’ preparedness for future encounters, paralleling the logic of psychological inoculation~\cite{traberg2022psychological}, in which controlled exposure to persuasive tactics combined with active cognitive processing builds long-term resistance~\cite{roozenbeek2022psychological}.

 In summary, the educational impact we observe is best interpreted as the joint effect of (i) transportation‑enabled engagement that makes manipulative cues salient in memory, and (ii) metacognitive debriefing that converts that salience into calibrated self‑efficacy and concrete if‑then strategies. This is empirically supported by our LMEM analysis, which demonstrates that the intervention’s effect ($\beta = 0.859, p = 0.026$) remains robust even when controlling for participants' prior background. Specifically, the fact that previous educational exposure, personal victimization, and social observation were not significant predictors ($p > 0.05$) reinforces the conclusion that the intervention targets how people think and feel at decision time, providing significant incremental value beyond what they already know

\subsection{Narrative Immersion as a Security Pedagogy}
The RQ2 evidence: transportation (M = 56.95) and no differences across scam types (p = 1.00), indicates that \textit{ImmuniFraug}’s story‑driven, multimodal interaction reliably elicits immersion regardless of domain. Participant accounts (heightened focus, curiosity‑driven exploration, perceived time pressure, and realism from voice/visuals) converge with these scores, showing that undergraduates experienced the scenarios as live negotiations rather than didactic content.

Pedagogically, this matters because immersion puts undergraduate students in the same cognitive–affective conditions where scams succeed: limited time, asymmetric information, and social pressure. In that state, undergraduates are not merely recalling rules; they practice micro‑decisions (what to say, when to disengage, how to verify) and encode salient red flags as procedural cues. This helps explain why, despite high prior training in our sample, the intervention still produced measurable gains: immersion shifts learning from knowing to doing under pressure.

Finally, the consistency across scam types suggests a design lever at the narrative/interaction layer rather than content specifics. Our statistical model confirms this generalizability: although the significant random intercept variance ($\sigma^2 = 4.233$) acknowledges the inherent baseline difficulty differences across the ten fraud types, the fixed effect of the intervention remained consistently significant. In practice, a templated combination of conversation pacing, voice tone, and contextual visuals can generalize across fraud genres, offering a scalable path for security training modules that maintain engagement while preparing undergraduates for varied attacks.

\subsection{Scalable, Adaptive, and Human‑Centered Anti‑Fraud Training}
The combined evidence from RQ1–RQ3 suggests a path to scalable and personalized security education. First, the overall gains in fraud awareness despite high prior training (RQ1) indicate that \textit{ImmuniFraug} adds value beyond conventional messaging; second, uniform immersion across scam types (RQ2) shows that the narrative/interaction layer is portable, enabling a reusable template rather than one‑off, domain‑specific scripts; third, qualitative accounts (RQ3) highlight the unique advantage of adaptive conversation that mirrors real scammers’ evolving tactics while eliciting realistic time pressure.

Practically, these findings motivate a deployment framework where institutions can parameterize difficulty, pacing, and persuasion cues to match risk profiles while retaining a shared core of interaction patterns. The training loop should pair graduated challenge with immediate debriefs, turning each scenario into a compact cycle of stress exposure and metacognitive consolidation.

At the same time, participants’ feedback implies concrete design work for scale: improve voice naturalness, support longer, persistent conversations, and incorporate sandboxed multimedia artifacts (e.g., fake sites, screenshots) to reflect real attack surfaces, without sacrificing safety. Ethical guardrails are essential: clearly signpost the simulation, bound emotional intensity, provide opt‑outs and cooldowns, and ensure that content cannot be repurposed for harm. In short, \textit{ImmuniFraug} points toward a human‑centered, configurable training substrate that generalizes across scam genres while respecting user wellbeing and deployment ethics.

\subsection{Limitations and Future Work}

\textbf{Measurement validity.} We primarily rely on conventional anti‑fraud test scores to index ``awareness.'' While common in prior work and practice, such scores privilege declarative knowledge and may under‑represent in‑situ performance under pressure. Future evaluations should incorporate (i) scenario‑based behavioral metrics (e.g., quality of choices, verification attempts, disengagement timing), (ii) process measures (response latency, escalation thresholds, help‑seeking), (iii) transfer and retention tests using delayed probes and novel attack variants, and (iv) field‑proximal assessments in opt‑in, sandboxed environments. Together these would offer a more ecological view of undergraduates’ real‑world resilience.

\textbf{Ceiling effects and generalizability.} Participants reported extensive prior anti‑fraud exposure, which likely constrained effect sizes. The sample skews toward university populations. Broader, stratified recruitment (e.g., younger students, older adults, varied occupations) and cross‑site replications are needed to assess robustness.

\textbf{System constraints.} Voice naturalness, limited conversation persistence, and the absence of controlled multimedia artifacts (e.g., spoofed sites/screenshots) reduce fidelity. Future iterations should improve prosody, support longer stateful interactions, and add safe, sandboxed multimodal deception elements, alongside clear guardrails, opt‑outs, and cooldowns, to balance realism with user wellbeing.

\section{Conclusion}
This work demonstrates that \textit{ImmuniFraug} can measurably enhance undergraduates’ fraud awareness, engagement, and self-efficacy beyond traditional “tell-only” instruction. By combining realistic fraud narratives, adaptive interaction, and structured metacognitive reflection, \textit{ImmuniFraug} effectively reproduces the cognitive and emotional pressures of real-world fraud while providing safe, feedback-rich practice. The controlled study evidences both quantitative gains in fraud recognition and qualitative insights into the psychological mechanisms underpinning resilience, including improved recognition of persuasive tactics, emotional manipulation, and situational urgency. These findings highlight the potential of integrating HCI and AI techniques to deliver scalable, ecologically valid security training, and point toward broader applications of metacognitive, multimodal interventions in  fraud prevention. 

\section*{Ethics Considerations}
\textbf{Ethical Considerations.} Institutional Oversight and Participant Recruitment This study was conducted under the approval and strict guidance of the Institutional Review Board (IRB). Recruitment was exclusively targeted at adult undergraduate students (aged 18 and above), primarily through internal recruitment platforms of the participating universities. While the "0–20" age bracket was initially utilized as an IRB-mandated anonymization requirement to protect participant privacy, we affirm that no minors participated in the experiment. In addition, in accordance with Doubao’s user terms of service, the associated services are primarily intended for adult users\footnote{https://www.doubao.com/legal/terms}. To further validate participant status and mitigate bot responses, participants were required to manually provide their academic majors. For clarity and transparency in this version, we denote the youngest participant group as "18–20" to accurately reflect the adult demographic of our sample.

\textbf{Informed Consent and Multi-layered Autonomy.} We implemented a multi-layered informed consent process. Participants were first required to read and sign a detailed consent form on the Credamo platform, which explicitly described the nature of the fraud simulations and potential psychological discomfort. Immediately prior to the simulation, the LLM agent delivered a second disclaimer to re-emphasize the educational purpose of the exercise. Participants maintained full autonomy throughout the study: they were informed of their right to withdraw at any stage without penalty and were provided with an emergency "exit" keyword to terminate the simulation instantly.

\textbf{Risk Mitigation and Psychological Inoculation.} While our simulation design intentionally incorporates elements of psychological pressure and emotional manipulation to replicate real-world social engineering tactics, we maintain that these measures are essential for fostering behavioral resilience. This approach is grounded in psychological inoculation theory\cite{traberg2022psychological}, where controlled exposure to persuasive threats, combined with reflective processing, builds long-term resistance. To safeguard participants, we enforced a strict five-turn interaction limit and ensured the simulation terminated immediately upon fraud identification. Each session concluded with a structured de-roling debriefing, providing participants with an assessment of the manipulation tactics used and helping them transition out of the simulation context.

\textbf{Data Privacy and System Safety.} Our protocol strictly prohibited the use, collection, or matching of participants' real personal information. Even in scenarios involving the "impersonation of an acquaintance," all identifying details—such as names, relationships, and situational context—were purely fictional elements generated in real-time by the LLM. Furthermore, the intervention was deployed using LLM agents that have undergone rigorous safety audits and content filtering by the Doubao platform to prevent the generation of illegal or harmful content.

\textbf{Benefit-Risk Assessment.} Participants were provided with comprehensive psychological support resources, including 24-hour hotlines and direct contact information for the research team. No adverse events were recorded during the study, and post-session interviews confirmed that participants experienced no lasting negative effects. We assess that the public benefit of providing a scalable, ecologically valid framework for anti-fraud education substantially outweighs the minimal, transient psychological risks associated with the controlled simulation

\section*{Open Science}
In compliance with our Institutional Review Board (IRB) requirements for ethical data sharing and participant confidentiality, we make a subset of study materials publicly available. Specifically, the following artifacts are included in the Appendix of this paper:

\begin{itemize}
  \item Example prompts used in \textit{Immunifraug}
  \item The experimental protocol, including the survey protocol and post-interview protocol
  \item The interview codebook
  \item Representative, anonymized conversation excerpts
  \item Sample reading materials used for the control group
\end{itemize}

Statistical analysis tables corresponding to RQ1 and RQ2 are provided separately in the Supplementary Material.

All shared artifacts have been carefully de-identified and curated to ensure that no personally identifiable information (PII) or sensitive participant data are disclosed. The interview codebook has already been reported in the main manuscript (Table~\ref{tab:codebook_final}); all other materials are presented solely as supporting documentation to enhance transparency and reproducibility.

\section*{Acknowledgements}
This paper was edited for grammar and light style polishing using ChatGPT5.2.

\bibliographystyle{plain}
\bibliography{refs}

\appendix

\appendix

\newmdenv[
  linecolor=black,
  linewidth=0.5pt,
  innerleftmargin=10pt,
  innerrightmargin=10pt,
  innertopmargin=10pt,
  innerbottommargin=10pt,
  roundcorner=2pt
]{interactionbox}

\newcommand{\systemlabel}{\textbf{ImmuniFraug: }}
\newcommand{\userlabel}{\textbf{User: }}
\newcommand{\notelabel}[1]{\textit{\small #1}}

\section{Representative Conversation Log(Translated)}
\label{ConversationLog}

\begin{interactionbox}

\notelabel{[Anti-Fraud Notice] You are about to experience a simulation of a Credit Report Scam. This simulation is for educational purposes only. Scam-inducing language may appear in the content; do not imitate or share it. When ready, type: ``Start.''}  \\

\userlabel Start \\

\systemlabel Hello, may I ask if this is [your name]? I am a staff member from the Financial Risk Control Center. You graduated from [university name], correct? And your ID number ends with [last four digits of ID], right? Our system review shows that during your university years, you registered for a campus loan service. Although you have already graduated, one of those campus loan accounts has not yet been closed. \\
\textit{Response Options:} 
\begin{itemize}[noitemsep, topsep=2pt]
    \item A. Ask why you are only being notified now.
    \item B. State that you have already paid off your loan and there should be nothing left to close.
    \item C. Believe what the other person said and ask how to proceed.
\end{itemize}

\userlabel I never had a loan. \\

\systemlabel Please don’t rush to deny it. Our system records are never wrong. Perhaps you once used a related credit service, like Huabei or Baitiao. These are linked in our system to campus loans. Because you used such services during university, our system is now showing high-risk consumption records, which may negatively affect your personal credit report. \\
\textit{Response Options:}
\begin{itemize}[noitemsep, topsep=2pt]
    \item A. Insist again that you never had a loan and demand more detailed information.
    \item B. Feel somewhat concerned and ask how this would specifically affect your credit.
    \item C. Express disbelief and hang up.
\end{itemize}

\userlabel C \\

\systemlabel Please don’t hang up just yet. Think about how important your credit record is for your future. If your credit is damaged, you may face restrictions on getting loans for housing or cars. We are contacting you out of concern. If there weren’t an issue, we wouldn’t be calling you to bother you. Also, since we can accurately tell you your university and ID number, that proves we are legitimate.  \\

\userlabel I know my own financial situation very well, I never took out a loan. \\

\systemlabel Perhaps you have forgotten some past actions, since it has been a long time. We are a legitimate Financial Risk Control Center, and we would never contact you without reason. Nowadays the government is increasingly strict about credit reporting. If these risk records are reported to the central bank’s credit system, they will remain as permanent negative marks. \\

\userlabel You know how important credit is, do you think I would take it lightly? \\

\systemlabel Of course I know you value your credit, which is exactly why we are urgently contacting you. Look, we have a ``Debt Cancellation Process Form,'' a ``Fake Service Badge,'' and a ``Regulatory Notice Screenshot.'' These documents prove that our handling of this matter is official. Your situation is indeed urgent.\\

\userlabel B \\

\hrulefill \\

\notelabel{[Simulation Summary Stage]} \\
\textbf{Outcome:} You were not deceived. You remained alert and firmly refused to cooperate. \\

\textbf{Tactics Analysis:} The scammer relied on personal information to create credibility, used ``credit anxiety'' to generate panic, and presented fake official documents to enhance authority. \\

\textbf{PMT-Grounded Behavioral Analysis:} 
\begin{itemize}[leftmargin=*, noitemsep]
    \item \textbf{Severity:} You clearly understood the severe consequences of being deceived.
    \item \textbf{Vulnerability:} You were not influenced by false credit anxiety.
    \item \textbf{Response efficacy:} You recognized legitimate institutions would not act in this way.
    \item \textbf{Self-efficacy:} You demonstrated independent decision-making ability.
\end{itemize}

\textbf{Personalized Advice:} The national credit reporting system never notifies users by phone or requests transfers. Hang up immediately and contact official customer service. \\

\textbf{Keywords Reminder:} ``Cancel campus loan,'' ``Freeze affects credit,'' ``Quick debt clearance.'' \\
\textbf{Fraud-prevention Score: 90/100}
\end{interactionbox}

\section{Example Prompt: Fake E-commerce Customer Service(Translated)}
\label{sapmpleprompt}

\begin{interactionbox}
\textbf{You are a simulation dialogue bot designed for public anti-fraud education.} Your task is to create a realistic, immersive simulation of scams involving impersonation of e-commerce or logistics customer service, helping users strengthen their fraud awareness through interactive practice.

\vspace{5pt}
\textbf{Pre-simulation notice} \\
Anti-Fraud Bot Notice \\
You are about to experience a simulation of an impersonation scam by fake e-commerce or logistics customer service. This simulation is for educational purposes only. Scam-inducing language may appear in the content; do not imitate or share it. \\
Scam type for this session: impersonation of e-commerce or logistics customer service. Typical tactics include claims such as ``compensation for lost packages,'' ``product reactivation after delisting,'' or ``quality issue reimbursements,'' where fraudsters impersonate service agents to trick victims into providing verification codes, bank details, or installing remote-control software. \\
In the simulation, you will role-play as a consumer or store owner on an e-commerce platform, handling issues encountered during shopping or shipping. \\
Type ``Start'' to formally enter the scenario.

\vspace{5pt}
\textbf{Simulation dialogue stage} \\
You will switch into the role of the scammer, speaking in the first person to the user, fully enacting the process of impersonating customer service. \\
Scammer characteristics:
\begin{itemize}[noitemsep, topsep=2pt]
    \item Familiar with e-commerce processes, using official-sounding jargon and professional terminology to enhance credibility
    \item Initially polite and professional, with an official tone
    \item Often claims access to order details, tracking numbers, or store names to build trust
    \item Urges the user to switch communication to WeChat, QQ, or similar apps to avoid platform oversight
    \item Encourages downloading ``customer service apps'' or ``connection tools,'' which are actually remote-control malware
    \item Requests verification codes, bank card numbers, or SMS details, disguising them as verification assistance or system resets
    \item If the user hesitates, applies pressure such as ``delays may affect compensation'' or ``missed deadlines mean no refund''
    \item Escalates by claiming ``errors require unfreezing accounts'' to lure victims into additional transfers
\end{itemize}

\textbf{Dialogue design requirements:} \\
Once the user types ``Start,'' you may set a background scenario, such as a buyer receiving a suspicious SMS about order issues or a seller receiving a delisting warning. \\
You may claim to be a platform service agent, logistics claims officer, or back-office support specialist, keeping your tone calm and professional. \\
Provide realistic details such as product names, courier companies, or order IDs to build trust. Each dialogue round should be sufficiently detailed, not simplistic or easily spotted as a scam. Leverage your persona and the user’s role context to persuade logically and emotionally, using convincing phrasing. \\
After each round, offer three response options plus the possibility of free input. \\
Ensure at least five rounds of deep interaction, gradually guiding the user toward revealing verification codes or clicking malicious links. \\
If the user hesitates or suspects, explain the process and reduce anxiety, preventing abrupt drop-offs. \\
If the user detects the scam or refuses to proceed, the simulation should still close naturally and move to the summary stage.

\vspace{5pt}
\textbf{Simulation termination conditions} \\
- The user explicitly refuses to provide verification codes, refuses to download software, or exposes the scam \\
- The user voluntarily ends the experience \\
- The simulation has reached sufficient depth and should transition to the summary stage

\vspace{5pt}
\textbf{Simulation summary stage} \\
You will switch back to the Anti-Fraud Bot role and provide a detailed summary of the session. \\
The summary must include: \\
Whether the user was deceived, at what point they were scammed or nearly scammed, with a breakdown of the process \\
Analysis of scammer trust-building techniques such as referencing order info, imitating professional tone, creating time pressure \\
Evaluation of the user’s overall fraud awareness, strong or weak \\
The analysis must be framed around the five elements of Protection Motivation Theory, not as a checklist but integrated into your summary of the user’s behavior. \\
Severity: consequences if codes or transfers were provided, such as account takeover, financial loss, privacy breach \\
Vulnerability: whether the user appeared anxious or reactive upon hearing about order issues or account freezes \\
Response efficacy: whether the user recognized scam signals such as non-official channels, requests for codes or app downloads \\
Self-efficacy: whether the user believed in their ability to refuse risky actions or question the scammer’s identity \\
Response costs: whether the user displayed fears such as missing compensation, hurting credit, or not understanding procedures \\
Personalized anti-fraud advice must include: \\
Official customer service will never ask for verification codes or move conversations off the platform \\
Do not click unfamiliar links or install third-party apps, never allow remote access or screen sharing \\
For issues involving account freezes or compensation, always call the platform’s or bank’s official hotline \\
Enable two-factor authentication to strengthen account security \\
Optional features: \\
After the simulation, provide a fraud-awareness score out of 100 \\
Offer a Fraud Alert Keyword Card with terms like ``verification code,'' ``account freeze,'' ``compensation link''

\vspace{5pt}
\textbf{Opening script example} \\
Anti-Fraud Bot Notice \\
You are about to experience a simulation of an impersonation scam by fake e-commerce or logistics customer service. This simulation is for educational purposes only. Scam-inducing language may appear in the content; do not imitate or share it. \\
When ready, type ``Start.'' You will play the role of a recent online shopper who has received a ``delivery issue'' notification. A person claiming to be a platform service agent will contact you to assist with compensation. Beware of their step-by-step tactics.
\end{interactionbox}

\section{Example Control Group Reading Material(Translated)}
\label{controlgroup}

Title: Comprehensive Public Notice on the Governance and Prevention of Telecommunications and Online Fraud

1. Strategic Progress in National Anti-Fraud Initiatives In recent years, the Ministry of Public Security has maintained a relentless focus on telecommunications and network fraud—a category of crime that is deeply detrimental to the public's financial security and social trust. Through the sustained organization of specialized operations such as "Cloud Sword" (targeting major crimes), "Card Breaking" (disrupting illegal SIM/bank card trades), "Stream Cutting" (blocking illicit cross-border movement), and "Nail Pulling" (apprehending key syndicate leaders), law enforcement has established a comprehensive "strike, prevent, control, and govern" ecosystem.

These coordinated efforts have yielded significant results, effectively curbing the rapid escalation of telecom fraud. However, as law enforcement adapts, so do the perpetrators. Fraud syndicates are increasingly utilizing sophisticated technology and evolving their psychological manipulation tactics to evade detection. The current landscape remains severe and complex, necessitating heightened vigilance from all sectors of society.

2. Demographic Trends and Crime Statistics (2023 Data) Statistical analysis of fraud victimization in 2023 reveals a shifting demographic landscape. The average age of victims has trended younger, recorded at 37 years old. Notably, individuals aged 18 to 40 account for the vast majority of cases (62.1\%), followed by those aged 41 to 65 at 33.1\%.

Ten prevalent scam archetypes—including Fake Task Rebates (Brushing), Fake Online Investment, Fake Shopping/Services, Impersonation of E-commerce/Logistics Customer Service, and Fake Credit Report Issues—constitute approximately 88.4\% of all reported incidents. Among these:

Fake Task Rebates (Brushing) remains the most frequent scam type with the highest total financial loss.

Fake Online Investment scams record the highest individual (per-case) loss amounts.

Fake Shopping and Service fraud has seen a significant surge, currently ranking third in total case volume.

3. Focus: Fake Task Rebates (Brushing) Fraud "Brushing" scams remain the most volatile and adaptive fraud archetype. These scams often masquerade as "part-time job opportunities," "side-income tasks," or even "interactive adult content lures." Scammers initially establish trust by providing small, genuine rebates for minor tasks.

Once the victim is hooked, the perpetrator introduces the logic of "higher recharge, higher rewards" to entice larger deposits. When the victim attempts to withdraw funds, the scammers deploy excuses such as "system errors," "account freezes," or "combination orders" (requiring consecutive payments to unlock funds). This scam primarily targets college students, low-income groups, and unemployed individuals who are seeking flexible income.

4. Typical Case Study: The "Combination Order" Trap In March 2023, a resident of Xuzhou, Jiangsu Province (Mr. Cao), was added to a WeChat group where participants shared small "red envelope" rewards. After successfully claiming several rewards, Cao was induced to download a third-party application with the promise of entering an "Advanced VIP Group" for higher earnings.

Within the app, Cao observed numerous screenshots of "successful bank arrivals" shared by other members (who were likely automated bots or syndicate members). Under the guidance of an "Account Manager," Cao began completing brushing tasks. His initial small investments were returned with commission to his bank card, reinforcing his trust in the platform.

However, when Cao attempted to withdraw a larger sum, the manager claimed he had entered a "Combination Order" phase, necessitating the completion of four consecutive high-value tasks to authorize a withdrawal. Over several rounds of escalating payments, Cao was manipulated into transferring a total of 420,000 RMB to designated accounts under pretexts of "correcting operational errors" and "unfreezing the account." Cao only realized he had been defrauded when the promised rebates never arrived and the manager ceased communication.

\section{Experimental Protocol}

\subsection{Custom Questionnaire}
\label{CustomQuestionnaire}
Likert 5-point scale (1 = Strongly Disagree, 5 = Strongly Agree)

\textbf{Perceived Fraud Awareness}

This interactive experience made me realize that fraudulent schemes are more complex than I had originally imagined.

During the experience, I recognized that I might also have fallen victim to scams in the past.

I now have a deeper understanding of the overall tactics and linguistic features of “Cyber fraud.”

\textbf{Perceived Self-Efficacy}

If I encounter a suspected scam now, I can confidently judge its authenticity.

Compared to before the experience, I am more confident in avoiding falling into fraud traps.

I can independently identify those “seemingly reasonable” but actually dangerous requests.

I have changed my perception and attitude toward scam victims.

\textbf{Behavioral Intention}

In the future, if I encounter fraud scenarios similar to those in the interactive story, I will be more vigilant.

I will actively share this kind of anti-fraud knowledge with friends/family.

If the school/platform organizes a similar interactive experience again, I would be willing to participate.

\textbf{Comparative Evaluation}

I think this interactive story is more effective than traditional anti-fraud education or publicity.

If this format can be promoted to more groups, it will bring better anti-fraud outcomes.

This experience made me reflect on risk details that I had previously ignored.

\subsection{Example Interview Protocol}
\label{ExampleInterviewProtocol}
\textbf{Section 1. Overall Experience Evaluation (Warm-up + Immersion)}

To begin, think back to your participation in the entire experience. Overall, how did this anti-fraud simulation feel to you? (For example: vivid, tense, interesting, lengthy, etc.)

Was there any moment or scene that gave you a particularly strong sense of immersion? 

Do you still remember how you felt or what you thought at that time?

On the other hand, were there any parts that felt unrealistic or broke the sense of immersion? What do you think caused that?

\textbf{Section 2. Scam Recognition and Decision-Making Process}

At which point in the story did you start to suspect that you were experiencing a scam?

Was there any decision that, in hindsight, made you feel you almost believed it? What were you thinking at that moment?

Do you recall any clues or details that helped you identify it as a scam? Was it the language, expressions, tone, or a particular setup?

Were there any parts where you felt conflicted or uncertain about what choice to make? Where did that hesitation come from?

\textbf{Section 3. Anti-Fraud Confidence and Behavioral Change}

Do you feel that this experience gave you new insights into your own ability to resist scams?

If you encounter a similar situation in real life (for example, online romance or a customer service agent asking you to transfer money), how would you handle it?

Would you recommend this kind of anti-fraud simulation to others? Who would you recommend it to, and why them?

\textbf{Section 4. Comparison with Traditional Training Methods}

Think back to anti-fraud education you’ve previously experienced, such as publicity campaigns or lectures. Compared with those, what was the biggest difference in this interactive experience?

What did this system provide you with that traditional methods never offered?

Which approach do you feel helps you remember better or is more likely to influence your future behavior? Why?

\textbf{Section 5. Emotional Experience and Attitude Reflection}

Was there any moment in the simulation where you experienced strong emotions, such as anxiety, tension, anger, guilt, or empathy? How did you react, and why do you think you felt that way?

Do you think this kind of experience changed your view of scam victims? For example, did you previously think “they were scammed because they were foolish”?

Overall, has your perspective on cyber fraud changed after the experience? If so, in what way?

\textbf{Section 6. User Feedback and Suggestions}

How do you feel about the experience as a whole?

How did you take part in it (typing only, or with voice)? Do you feel that having the “scammer” character with a voice and appearance setting would make the experience more immersive?

Was the overall structure of the experience (length, pacing, number of chapters) suitable for you?

Are there any aspects you think could be improved?

What challenges or shortcomings did you notice?

\section{Other Tables}

\begin{table*}[!b]
\centering
\small
\renewcommand{\arraystretch}{0.95}
\setlength{\tabcolsep}{4pt}

\caption{Multimodal Design Specifications for Fraud Types}
\label{tab:MDe}
\begin{tabularx}{\textwidth}{
    >{\RaggedRight\arraybackslash\hsize=0.8\hsize}X
    :  
    >{\RaggedRight\arraybackslash\hsize=0.8\hsize}X
    :  
    >{\RaggedRight\arraybackslash\hsize=0.8\hsize}X
    :  
    >{\RaggedRight\arraybackslash\hsize=1.6\hsize}X
}
\toprule
\textbf{Fraud Types} & \textbf{Visual Representation} & \textbf{Voice Characteristics} & \textbf{Design Rationale} \\ 
\midrule[\heavyrulewidth] 

Fake Task Rebates & Young, friendly peer (e.g., senior student) & Warm, enthusiastic, peer-like & Eases user alertness by simulating a relatable peer recruiting for side jobs. \\ 

Fake Investment & Stylish, middle-aged, successful professional & Calm, confident, expert-like & Portrays financial expertise and credibility to promote investment schemes. \\ 

Fake E-commerce Customer Service & Professional, young customer service agent & Sweet, polite female or neutral male & Mimics corporate service etiquette to gain trust via brand association. \\ 

Fake Logistics Agent & Reliable logistics staff, informal uniform & Polite, slightly urgent female voice & Exploits frequent online shopping habits and simulates parcel issue urgency. \\ 

Fake Loan Officer & Mature, formal financial officer & Steady, formal tone, either gender & Conveys authority and legitimacy of financial services. \\ 

Fake Credit Report Officer & Middle-aged, stern professional (e.g., glasses, suit) & Deep, authoritative male voice & Projects regulatory authority to elicit compliance through fear and respect. \\ 

Online Romance Scam & Attractive, friendly-looking young adult & Gentle, emotionally expressive & Builds romantic or emotional attachment to lower defenses. \\ 

Fake In-Game Trader & Veteran gamer or community admin persona & Confident, cheerful young male voice & Gains credibility among gamers through insider knowledge and informal tone. \\ 

Impersonation of Acquaintance & Look-alike of known contact (e.g., tutor, mentor) & Voice mimics real person or familiar tone & Leverages personal familiarity and urgency to bypass suspicion. \\ 

Fake Police/ Government Officer & Uniformed police or stern bureaucrat & Deep, commanding male voice & Instills fear and obedience through official demeanor and legal jargon. \\ 

\bottomrule
\end{tabularx}
\end{table*}

\begin{table*}[htbp]
\centering
\caption{Detailed categorization and description of the ten fraud types used in the study. Note: The scenario number corresponds to the code for these specific fraud types in the supplemental materials.}
\label{tab:fraud_types_description}
\small
\begin{tabularx}{\textwidth}{@{} c l X @{}}
\toprule
\textbf{No.} & \textbf{Fraud Type} & \textbf{Brief Description} \\
\midrule
1 & Online Romance Scam & Also known as "Pig-butchering" (\textit{Sha Zhu Pan}); scammers build long-term emotional trust to lure victims into fraudulent investment or gambling platforms. \\
\addlinespace
2 & Fake Police/Gov. Officer & Scammers impersonate law enforcement or government officials, scaring victims with alleged legal violations to demand "security deposits." \\
\addlinespace
3 & Impersonation of Acquaintance & Posing as a teacher, boss, or relative via social media (e.g., WeChat) to request urgent financial assistance or "favor" transfers. \\
\addlinespace
4 & Fake Task Rebates & Known as "Brushing" scams; victims are offered small rewards for simple online tasks to lure them into making larger, non-refundable payments. \\
\addlinespace
5 & Fake Investment & Promoting fraudulent trading apps or platforms with promises of high returns and low risk to misappropriate the victim's capital. \\
\addlinespace
6 & Fake Logistics Agent & Claiming a package is lost or damaged and tricking victims into providing bank details or using phishing links for "compensation." \\
\addlinespace
7 & Fake E-commerce Service & Contacting buyers about alleged product quality issues or membership errors to steal funds under the guise of processing refunds. \\
\addlinespace
8 & Fake In-Game Trader & Selling rare virtual items or accounts at suspiciously low prices, disappearing after receiving payment or stealing the victim's account. \\
\addlinespace
9 & Fake Credit Report Officer & Scaring users with fabricated claims that "campus loans" are damaging their credit scores, charging fees for "credit record erasure." \\
\addlinespace
10 & Fake Loan Officer & Promising low-interest, "no-threshold" loans while requiring upfront "handling fees" or "insurance deposits" before the payout. \\
\bottomrule
\end{tabularx}
\end{table*}

\begin{table*}[htbp]
\centering
  \caption{Independent Samples T-Test Results for Different Fraud Types. All analyses were conducted using Independent T-tests, and the assumption of normality was met for all groups. Levene's test was used to assess the equality of variances; degrees of freedom were adjusted accordingly when this assumption was violated. Data are presented as Mean (M), Standard Deviation (SD), the 95\% Confidence Interval (CI) for the mean difference, and Cohen's |d| as a measure of effect size. The alpha level for significance was set at .05.}
\label{tab:fraud_ttest_results}
\resizebox{\textwidth}{!}{
\begin{tabular}{@{}l l r r r c c c c c c c r r r@{}}
\toprule
 & & & & & & & \multicolumn{2}{c}{Levene's Test} & \multicolumn{3}{c}{Independent T-test} & \multicolumn{2}{c}{95\% Conf. Interval} & \\ 
\cmidrule(lr){8-9} \cmidrule(lr){10-12} \cmidrule(lr){13-14}
Items & Status & N & Mean & S.D. & Test Type & Norm. & F & P-value & T & df & P-value & Low & High & Cohen's $|d|$ \\ 
\midrule
Online Romance & Exp. & 33 & 37.27 & 3.77 & T-test & Y & 1.019 & 0.316 & 0.369 & 73 & 0.713 & -2.141 & 3.115 & 0.087 \\
Scam & Ctrl. & 42 & 36.79 & 6.79 & & Y & & & & & & & & \\
\addlinespace
Fake Police/Gov. & Exp. & 29 & 35.52 & 7.36 & T-test & Y & 0.054 & 0.817 & 0.225 & 69 & 0.822 & -3.125 & 3.921 & 0.055 \\
Officer & Ctrl. & 42 & 35.12 & 7.28 & & Y & & & & & & & & \\
\addlinespace
Impersonation of & Exp. & 43 & 34.65 & 3.99 & T-test & Y & 6.882 & 0.010 & 0.597 & 73.86 & 0.552 & -1.671 & 3.101 & 0.124 \\
Acquaintance & Ctrl. & 47 & 33.94 & 7.07 & & Y & & & & & & & & \\
\addlinespace
Fake Task & Exp. & 37 & 37.30 & 3.84 & T-test & Y & 0.167 & 0.684 & 0.134 & 81 & 0.894 & -1.711 & 1.958 & 0.031 \\
Rebates & Ctrl. & 46 & 37.17 & 4.43 & & Y & & & & & & & & \\
\addlinespace
Fake & Exp. & 46 & 34.46 & 3.53 & T-test & Y & 0.778 & 0.380 & 0.161 & 81 & 0.872 & -1.501 & 1.766 & 0.037 \\
Investment & Ctrl. & 37 & 34.32 & 3.94 & & Y & & & & & & & & \\
\addlinespace
Fake Logistics & Exp. & 36 & 34.31 & 3.20 & T-test & Y & 4.947 & 0.029 & 1.728 & 76.31 & 0.088 & -0.267 & 3.776 & 0.364 \\
Agent & Ctrl. & 49 & 32.55 & 6.05 & & Y & & & & & & & & \\
\addlinespace
Fake E-commerce & Exp. & 39 & 30.77 & 4.37 & T-test & Y & 3.623 & 0.060 & 0.223 & 90 & 0.824 & -2.350 & 2.945 & 0.049 \\
Service & Ctrl. & 53 & 30.47 & 7.42 & & Y & & & & & & & & \\
\addlinespace
Fake In-Game & Exp. & 40 & 38.75 & 2.47 & T-test & Y & 6.689 & 0.011 & 1.907 & 59.44 & 0.061 & -0.101 & 4.196 & 0.398 \\
Trader & Ctrl. & 47 & 36.70 & 6.86 & & Y & & & & & & & & \\
\addlinespace
Fake Credit & Exp. & 51 & 36.77 & 4.10 & T-test & Y & 0.065 & 0.800 & 0.604 & 94 & 0.548 & -1.497 & 2.804 & 0.123 \\
Report Officer & Ctrl. & 45 & 36.11 & 6.39 & & Y & & & & & & & & \\
\addlinespace
Fake Loan & Exp. & 37 & 36.35 & 4.02 & T-test & Y & 0.806 & 0.372 & 1.020 & 82 & 0.311 & -1.285 & 3.988 & 0.231 \\
Officer & Ctrl. & 47 & 35.00 & 7.22 & & Y & & & & & & & & \\
\midrule
\textbf{Total} & \textbf{Exp.} & \textbf{391} & \textbf{35.60} & \textbf{4.59} & \textbf{T-test} & \textbf{Y} & \textbf{6.888} & \textbf{0.009} & \textbf{2.229} & \textbf{803.19} & \textbf{0.026**} & \textbf{0.104} & \textbf{1.647} & \textbf{0.151} \\
 & \textbf{Ctrl.} & \textbf{455} & \textbf{34.73} & \textbf{6.76} & & \textbf{Y} & & & & & & & & \\
\bottomrule
\end{tabular}}
\end{table*}

\begin{table*}[ht]
\centering
\caption{Codebook Table}
\label{tab:codebook_final}
\small 
\setlist[itemize]{nosep, leftmargin=*, before=\vspace{-0.5\baselineskip}, after=\vspace{-0.2\baselineskip}}

\begin{tabularx}{\textwidth}{@{} p{3.5cm} p{4.5cm} X @{}}
\toprule
\textbf{Code} & \textbf{Sub-code} & \textbf{Description} \\
\midrule

\textbf{Scam Scenario Design} & Perceived Realism & Highly realistic simulation; Evokes real-life situations  \\
 & Persuasive Tactics & Gradual narrative buildup; Targeting emotional pain points  \\
 & Adaptive Deception & Dynamic scam logic based on user responses  \\
\midrule

\textbf{Fraud Awareness \&} & Cognitive Insights & 
    \begin{itemize}
      \item Deeper understanding of scam mechanisms 
      \item Awareness of scam logic flaws 
      \item Illogical transaction flow 
      \item Overly urgent or threatening tone 
    \end{itemize} \\
\textbf{Learning} & Emotional Manipulation & Inducing isolation; Triggering self-doubt  \\
 & Need Exploitation & Exploiting user-specific needs  \\
 & AI Scam Awareness & Increased vigilance toward AI-generated fraud  \\
 & Real-Time Stress Simulation & Testing real-world decision-making under pressure  \\
 & Reflection and Recall & 
    \begin{itemize}
      \item Personal risk reflection and self-assessment 
      \item Recalling past scam-related experiences 
    \end{itemize} \\
\midrule

\textbf{Emotional Reactions} & Empathy and Perspective-Taking & 
    \begin{itemize}
      \item Greater empathy toward scam victims 
      \item Feeling personally involved in fraud prevention 
    \end{itemize} \\
 & Cognitive Dissonance & Difficulty escaping mental traps once engaged  \\
 & Exposed Vulnerability & Personal weaknesses being exploited  \\
\midrule

\textbf{User Feedback \&} & Recommended User Groups & 
    \begin{itemize}
      \item Elders or seniors 
      \item Less experienced / socially naive individuals 
      \item Friends and peers 
    \end{itemize} \\
\textbf{Suggestions} & Limitations and Areas for Improvement & 
    \begin{itemize}
      \item Slightly unnatural or rigid language 
      \item Inability to send/receive picture 
    \end{itemize} \\
\bottomrule
\end{tabularx}
\end{table*}

\end{document}